\documentclass[12pt,a4paper]{article}
\usepackage{amsfonts}
\usepackage[table, dvipsnames]{xcolor}

\usepackage{amsmath}
\usepackage{array}
\usepackage{arydshln}
\usepackage{rotating}
\usepackage{amssymb}
\usepackage{bm}
\usepackage{amsmath}
\usepackage{geometry}
\usepackage{setspace}
\usepackage{graphicx}
\usepackage{lscape}
\usepackage{verbatim}

\usepackage[authoryear]{natbib}
\setcitestyle{authoryear, open={(},close={)}}

\usepackage[table]{xcolor}
\usepackage{calrsfs}
\usepackage{mathrsfs}
\usepackage{algorithm2e}
\usepackage{algpseudocode}
\usepackage{tablefootnote}
\usepackage{tabularx}
\usepackage{ragged2e}
\definecolor{winered}{rgb}{0.5,0,0}
\definecolor{LightCyan}{rgb}{0.88,1,1}
\usepackage[bookmarks=true, bookmarksnumbered=true, allbordercolors={1 1 1}]{hyperref}
\hypersetup{
  colorlinks   = true, 
  urlcolor     = blue, 
  linkcolor    = blue, 
  citecolor    = winered,
}
\usepackage[open=true, numbered=true]{bookmark}
\usepackage{float}
\usepackage{fancyhdr}
\usepackage[toc,page]{appendix}
\usepackage{scalefnt}
\usepackage{afterpage}
\usepackage[left]{lineno}
\usepackage{caption}
\usepackage{subcaption}
\usepackage{varioref}
\usepackage{cleveref}
\usepackage{booktabs}
\usepackage{dsfont}
\usepackage{siunitx}

\makeatletter
\def\blfootnote{\xdef\@thefnmark{}\@footnotetext}
\makeatother

\emergencystretch=\maxdimen
\newcommand{\indep}{\perp \!\!\! \perp}

\begin{document}
	
\title{\textbf{Bayesian modelling of VAR precision matrices using stochastic block networks}\blfootnote{We would like to thank Luca Barbaglia, Sune Karlsson, James Mitchell, Luca Onorante, Michael Smith, Mike West as well as participants of the 13th European Seminar on Bayesian Econometrics (Glasgow, 2023), the yearly meeting of the Austrian Economic Association 2023 (Salzburg, 2023), the 6th Annual Workshop on Financial Econometrics (Örebro, 2023) and the 17th International Conference on Computational and Financial Econometrics (Berlin, 2023) for helfpul comments and suggestions. Huber and Scheckel gratefully acknowledge funding  from the Austrian Science Fund (FWF, grant no. ZK-35) and the Jubiläumsfond of the Oesterreischische Nationalbank (OeNB, grant no. JF-18740 and JF-18763).
}}


\author{
    \begin{tabular}{rl c rl}
         Florian & HUBER & & Gary & KOOP  \\
         \multicolumn{2}{c}{\small University of Salzburg} & & \multicolumn{2}{c}{\small University of Strathclyde} \\
         \multicolumn{5}{c}{}\\
         Massimiliano & MARCELLINO & & Tobias & SCHECKEL \\
         \multicolumn{2}{c}{\small Bocconi University} & & \multicolumn{2}{c}{\small University of Salzburg} \\
         \multicolumn{2}{c}{\small CEPR, IGEIR, BAFFI, BIDSA} & & \multicolumn{2}{c}{\small WIFO}\\
    \end{tabular}
}
\vspace{3cm}
\date{\today}

\maketitle

\vspace{1mm}
\vspace{1mm}
\vspace{-0.25cm}
\begin{abstract}
\noindent Commonly used priors for Vector Autoregressions (VARs) induce shrinkage on the autoregressive coefficients. Introducing shrinkage on the error covariance matrix is sometimes done but, in the vast majority of cases, without considering the network structure of the shocks and by placing the prior on the lower Cholesky factor of the precision matrix. In this paper, we propose a prior on the VAR error precision matrix directly.  Our prior, which resembles a standard spike and slab prior,  models variable inclusion probabilities through a stochastic block model that clusters shocks into groups. Within groups, the probability of having relations across group members is higher (inducing less sparsity) whereas relations across groups imply a lower probability that members of each group are conditionally related. We show in simulations that our approach recovers the true network structure well. Using a US macroeconomic data set, we illustrate how our approach can be used to cluster shocks together and that this feature leads to improved density forecasts. 
\end{abstract}

\thispagestyle{empty} 

\begin{flushleft}
    
\end{flushleft}


\clearpage

\setcounter{page}{1}
\doublespacing
\section{Introduction}
Vector Autoregressions (VARs) are routinely used for structural analysis and predictive inference in academia and policy institutions. These models, however, suffer from overfitting issues if the number of time series is large. Bayesian solutions rely on shrinkage priors so as to force coefficients associated with irrelevant predictors towards zero and thus improve inference \citep[see, among many others,][]{carriero_etal2009, banbura_etal2010, koop2013, korobilis2013, giannone_etal2014, huber_feldkircher2019}.

These papers propose shrinking the VAR coefficients towards a known location  and by doing so reducing the number of effective variables. This makes sense in light of the huge number of regression coefficients. For moderately-sized panels one often overlooked problem is that the number of parameters in the covariance matrix grows quadratically in the number of time series. Except in a few papers working with noninformative priors in VARs of low dimension, Bayesians use informative priors such as the inverse-Wishart prior on the error covariance matrix $\bm \Sigma$ (or equivalently the Wishart prior on the error precision $\bm \Omega := \bm \Sigma^{-1}$) and, thus, induce shrinkage. However these priors do not induce parsimony (i.e. by restricting error covariances to be exactly or nearly zero).  

Some papers that induce shrinkage on the error covariance matrix  are  \cite{george_etal2008} and \cite{koop2013} which use  priors such as the stochastic search variable selection (SSVS) prior.  These priors assume that the variable selection parameters apply to each coefficient individually and independently (i.e. each error covariance is either shrunk to zero or not, independently of the other covariances). Thus, they do not capture situations where covariances cluster together. Moreover, these techniques introduce shrinkage on the lower Cholesky factor of $\bm \Sigma$ or $\bm \Omega$, implying that they are not order-invariant. In a recent paper,  \cite{arias_etal2023_je}, show that the (lack of) order invariance matters for predictive likelihoods and hence, having techniques that are order-invariant might be preferable.\footnote{For recent, order-invariant, approaches, see \cite{chan2024large} and \cite{huber2023fastorderinvariantinferencebayesian}.}

In this paper,  we develop a new method that takes the network structure of the contemporaneous relations into account and is order-invariant. Our model assumes a time-varying covariance matrix that decomposes the covariance matrix into a time-varying and a time-invariant part. The time-invariant part establishes the contemporaneous relations across equations. On this part, we introduce our new shrinkage prior. Our prior draws on insights from the network literature and, in particular, is based on a stochastic block model \citep[(SBM), see, e.g.,][]{legramanati_etal2022}. 

For many applications in macro and finance the covariance or precision matrices of the shocks can be interpreted as a network that encodes the contemporaneous relationships between different units (such as companies, countries or variable types) and these shocks might form clusters. For instance, shocks specific to a region might have an immediate effect only on variables within that given region whereas other regions are impacted only with a time lag through the VAR coefficients. These patterns have implications for the covariances of the shocks, leading to a possibly sparse structure. For example, if such a regional structure exists, a successful shrinkage prior should not introduce shrinkage on the blocks of the precision matrix within a given region  but should force blocks between regions to zero. Standard shrinkage priors including SSVS and variants thereof do not take the presence of clusters into account. \cite{ahelegbey_etal2016_jae, ahelegbey2016_aas, billio_etal2019} use graphical models to shrink covariance matrices. However, their priors on the network still assume that the edge probabilities are independent a priori. In addition, to sample the network they require a Metropolis Hastings (MH) step. This can suffer from poor mixing, especially since the parameter space is extremely large. 

This motivates the present paper. Our goal is to set up a standard VAR with heteroskedastic shocks. We use an SSVS prior \citep{george_mcculloch1993} to shrink the elements of the precision matrix to zero. However, instead of assuming that the indicators that control whether a given precision parameter should be forced to zero or not arise from a Bernoulli with a common (and often fixed) prior inclusion probability, we model the latter using a SBM \citep[see, e.g.,][]{holland_etal1983, nowicki_snijders2001, legramanati_etal2022}. The resulting model endogenously detects clusters and thus introduces shrinkage on the VAR precision matrix while taking possible within- and cross-cluster linkages into account. The SBM we use does not require the pre-selection of the number of clusters, rather it is inferred adaptively alongside the remaining model parameters.  

To make sure that our resulting SBM-VAR model is scalable to large dimensions, we develop an efficient  Markov Chain Monte Carlo (MCMC) sampler to simulate from the joint posterior distribution. This sampler builds on recent advances in MCMC estimation of large VARs \citep[see][]{carriero_etal2022}.

To evaluate and illustrate our approach, we first carry out a thorough simulation exercise. Using synthetic data generated from a variety of alternative specifications, we show that our model accurately recovers the true network structure, and generally improves upon the standard SSVS prior. 

We then move on to a forecasting exercise involving a large set of US economic variables. Similar to our synthetic data exercise, our model finds parsimonious network structures although it is interesting to note that these change somewhat over time both in terms of the number of clusters and their modularity. The forecasting exercise shows that this property tends to lead (with some exceptions) to  better forecast performance than either an SSVS prior which does not incorporate a network structure or a non-informative prior. 

The remainder of the paper is structured as follows. Section 2 introduces the basic VAR model, briefly describes commonly used priors on the VAR coefficients and the error covariances, provides a brief introduction to stochastic block models in the context of VAR error precision matrices, introduces the SBM-VAR, specifically focusing on the VAR precision prior, and discusses the MCMC sampler. Section 3 provides simulation evidence that our model works well. Section 4 presents the empirical applications. Section 5 summarizes and concludes the paper. An Appendix contains additional details on the sampler and on the data used in the empirical application.

\section{Stochastic Block Network VARs}
\subsection{VARs with Stochastic Volatility}
Our goal is to model an $M$-dimensional vector of time series $\bm y_t$ that is observed at time $t=1, \dots, T$ using a VAR model:\footnote{In our theoretical discussion we ignore deterministic terms in the model. In our empirical application we include an intercept.}
\begin{equation}\label{eq:var}
  \bm y_t= \bm A_1 \bm y_{t-1} + \dots + \bm A_P \bm y_{t-P} +  \bm \varepsilon_t,\quad \bm \varepsilon_t \sim \mathcal{N}(\bm 0, \bm \Sigma_t),
\end{equation}
where $\bm A_j ~ (j=1,\dots, M)$ are $M \times M$ coefficient matrices, $P$ is the maximum lag length and the shocks in $\bm \varepsilon_t$ are independent, zero mean and normally distributed with a time-varying  variance-covariance matrix $\bm \Sigma_t$. 

The existing literature has many treatments of the error covariance matrix for VARs based on different decompositions and the manner in which time-variation is allowed for. Perhaps the most common, see e.g. \cite{CS_2005}, is to take a Cholesky decomposition $\bm \Sigma_t = \bm L \bm D^2_t \bm L^{'}$ where $\bm L$ is lower triangular with ones on the diagonal and $\bm D^2_t = \text{diag}(e^{d_{1, t}}, \dots, e^{d_{M,t}})$ is a diagonal matrix of time-varying variances. It is common to assume, as we do in this paper, that $d_{j, t}$ evolves according to independent AR(1) models: $d_{j, t} = \rho_j d_{j, t-1} + \sigma_{\rho, j} v_{j,t}$ with $\rho_j$ denoting the persistence parameter and $\sigma^2_{\rho, j}$ the error variance.

Use of the Cholesky decomposition has been criticized in papers such as  \cite{arias_etal2023_je} since it leads to a lack of order invariance.\footnote{\cite{arias_etal2023_je} also discusses a range of order invariant approaches and discusses their properties. Our method for decomposing the error covariance matrix is not the same as the order invariant specification of \cite{AH_2021} but does share a similar structure to it. } In this paper, we use a different decomposition of $\bm \Sigma_t$ which has attractive properties for our purposes in terms of ease of interpretation and computational efficiency. 

To explain our specification, we begin with the following decomposition and define  $\bm \Omega_t := \bm \Sigma_t^{-1}$ which is the error precision matrix: 
\begin{equation*}
    \bm \Sigma_t = \bm D_t \bm \Sigma \bm D_t\quad \Leftrightarrow\quad  \bm \Omega_t = \bm D_t^{-1} \bm \Omega \bm D_t^{-1},
\end{equation*}
where $\bm D_t = \text{diag}(e^{d_{1, t}/2}, \dots, e^{d_{M,t}/2})$ is a diagonal matrix of time-varying standard deviations so that $\bm D_t^2 = \bm D_t \odot \bm D_t$ with $\odot$ meaning element-wise multiplication. For convenience, let $\bm d_t = (d_{1,t}, \dots, d_{M, t})'$ denote the $M-$dimensional vector of time-specific log-volatilities.  Note that the decomposition involves two time-invariant matrices, $\bm \Sigma$ and $\bm \Omega $ which, with some abuse of terminology we still refer to as covariance and precision matrices. That is, they are the covariance and precision matrices abstracting from the time-varying standard deviations. If elements of $\bm \Omega$ ($\bm \Sigma$) are zero then the corresponding elements of $\bm \Omega_t$ ($\bm \Sigma_t$) are zero. 

Note that we face a choice as to whether to place our SBM prior on $\bm \Omega $ or $\bm \Sigma$. Either is possible, but we place it on $\bm \Omega$. This is for reasons outlined in papers such as \cite{FLL_2016}. This paper points out that covariance and precision matrices contain different types of information. In the context of sparse estimation (such as we are attempting to do using our SBM prior), working with the precision matrix may be more sensible since precision matrices are often sparser than covariance matrices. A sparse precision matrix does not necessarily imply a sparse covariance matrix (and vice versa). \cite{FLL_2016} point out that the covariance matrix relates to marginal correlations between variables but the precision matrix relates to conditional correlations. That is, the precision matrix encodes the conditional dependence between $\bm \varepsilon_{i, \bullet}$ and $\bm \varepsilon_{j, \bullet}$ the $i$ and $j^{th}$ column of $\bm E = (\bm \varepsilon_1, \dots, \bm \varepsilon_T)'$. And these conditional correlations can be mapped into an undirected graph. In essence, if a sparse graphical or network structure exists it is better to model it via the precision matrix rather than the covariance matrix. 

Decomposing $\bm \Sigma_t$ into $\bm \Sigma_t = \bm D_t \bm \Sigma \bm D_t$ implies that the $(i, j)^{th}$ element in $\bm \Sigma_t$ is scaled by $e^{d_{i,t}/2} \cdot e^{d_{j,t}/2}$ and hence the covariances depend on the volatility dynamics across the different equations. Moreover, and this constitutes another advantage of this decomposition, if the prior is placed on $\bm \Omega$, then the model is order-invariant. It would not be so if we used some sort of SBM prior in the Cholesky-decomposed model.  

$\bm \Sigma$ can be further decomposed into $\bm \Sigma = \bm D \bm R  \bm D$ with $\bm D = \text{diag}(e^{d_1/2}, \dots, e^{d_M/2})$ and $\bm R$ denoting the correlation matrix. The matrix $\bm D^2 = \bm D \odot\bm D$ can be seen to introduce an intercept into the AR(1) models for the log-volatilities since $\bm D_t \odot \bm D = \text{diag}(e^{(d_1 + d_{1,t})/2}, \dots, e^{d_M + d_{M,t}/2})$. As we will discuss below, we use this further decomposition for computational reasons since, conditional on $\bm D_t$, the rows and columns in $\bm \Sigma$ can be sampled using a Gibbs sampler. Other related specifications, such as that of \cite{AH_2021}, which do not use this decomposition of $\bm \Sigma$ can be estimated using Bayesian MCMC methods. However, these MCMC algorithms are not straightforward Gibbs samplers and are not scalable to large dimensions.   

\subsection{Network-based priors for VARs}
If the number of variables $M$ (and/or the lag length $P$) is large, the model suffers from overfitting. Most notably, the number of dynamic coefficients in the VAR is $P \times M^2$ and thus grows quadratically with $M$. Bayesian shrinkage priors such as the Minnesota prior  \citep[e.g.,][]{doan_etal1984, littermann1986, giannone_etal2015}) or global-local shrinkage priors \citep[e.g.,][]{huber_feldkircher2019, follet_yu2019, kastner2020sparse} introduce shrinkage so as to reduce the effective number of elements in the VAR coefficients. However, as stated in the introduction, these priors either introduce little shrinkage on the $M (M-1)/2$ free elements of the error covariance matrix or do so in an unstructured manner.  In the homoskedastic case, it is common to use an inverse Wishart prior on the covariance matrix of the VAR (reduced form) errors \citep[e.g.,][]{kadiyala_karlsson1997, giannone_etal2015, chan2021}. Often relatively non-informative choices are made for the prior hyperparameters of the inverse Wishart prior. 
Many other papers, which allow for stochastic volatility, use a Gaussian prior for the lower triangular $\bm L$ (or $\bm L^{-1})$, see, e.g., \cite{CS_2005, carriero_etal2019}.  This strategy gives rise to three issues. First, it does not take a possible network structure into account. Second,  it is not clear what shrinkage on $\bm L$ implies for $\bm \Sigma_t$ or $\bm \Omega_t$. Third, placing a prior on $\bm L$ implies that all empirical findings depend on how the elements in $\bm y_t$ are ordered. The goal of this paper is to develop a shrinkage prior that surmounts these problems.  For the reasons given in the preceding sub-section, we will do so by constructing a prior designed to capture a sparse network structure on $\bm \Omega$ and, thus, the error precision matrix $\bm \Omega_t$.

Let $\delta_{i,j}$ denote a binary indicator that equals $1$ if there exists a relationship between $\varepsilon_{i,t}$ and $\varepsilon_{j,t}$ and $0$ if they are assumed to be conditionally independent. The indicator $\delta_{i,j}$ will be used to set up a shrinkage prior on the elements in $\bm \Omega$. Following \cite{george_mcculloch1993} and \cite{george_etal2008}, we assume that the off-diagonal elements of $\bm \Omega$, $\omega_{i,j}~(i=2, \dots, M; j=1, \dots, M-1; i < j)$ arise from a mixture of two Gaussian distributions:
\begin{equation}\label{eq:ssvs}
    \omega_{i,j} \sim \delta_{i,j}\mathcal{N}\left(0,  \tau_{ij, 1}^2) + (1-\delta_{i,j}) \mathcal{N}(0,  \tau_{ij, 0}^2\right),
\end{equation}
with $\tau_{ij, 1}^2$ (called the 'slab' variance) and $\tau_{ij,0}^2$ (called the 'spike' variance) denoting prior scaling parameters such that $\tau_{ij,1} \gg \tau_{ij,0}$. These scaling parameters can be set using an empirical Bayes approach which involves the use of the sample standard deviation of the time series under consideration or using the OLS estimate of it from an auxiliary regression.  Another approach is to follow \cite{ishwaran_rao2005} who let $\tau_{ij,0}^2 = c \cdot \tau_{i,j}^2$ with $c$ being a constant close to zero and $\tau_{i,j}^2$ is a common (over the two distributions in the mixture) scaling parameter.

Given that the OLS estimates of $\bm \Omega$ are difficult to obtain due to the presence of SV, we follow the latter approach and estimate the hyperparameters of the prior.  We follow \cite{ishwaran_rao2005} and place an inverse gamma prior on the slab variance, such that:
\begin{equation*}
    \tau_{i,j}^2 \sim \text{InvGa}(a_\tau, b_\tau),
\end{equation*}
where $\text{InvGa}(\cdot,\cdot)$ denotes the inverse gamma distribution.\footnote{Here, a random variable $z$ follows an inverse Gamma distribution with density $p(z|\alpha,\beta) = \frac{\beta^\alpha}{\Gamma(\alpha)}  z^{-\alpha-1} \text{exp}(-\frac{\beta}{z})$.}  We set $\alpha_\tau = 5$ and $\beta_\tau = 4$. We obtain the spike variance by multiplying $\tau^2_{i,j}$ with the constant $c = 2.5^{-5}$.

The key point to make here is that $\delta_{i,j}$ is a binary indicator that controls which component to use and prior independence is assumed over the elements of $\bm \Omega$. In other words, the following relationship holds:
\begin{equation}
    \delta_{i,j} = 0\quad  \Leftrightarrow \quad  \bm \varepsilon_{\bullet, i} \indep \bm \varepsilon_{\bullet, j}|\{ \bm \varepsilon_{\bullet, s}\}_{s \neq i,j} \quad\Leftrightarrow\quad\omega_{ij} \approx 0,
\end{equation}
implying that $\bm \varepsilon_{\bullet, i}$ and  $\bm \varepsilon_{\bullet, j}$ are approximately independent given the other shocks in the system. A typical assumption is that the indicators arise from a Bernoulli prior distribution with prior probability $\underline{\pi}$:
\begin{equation*}
    \delta_{i,j} \sim Bernoulli(\underline{\pi}).
\end{equation*}
Hence, the probability that $\delta_{i,j}=1$ is  $\underline{\pi}$ a priori which is the same for all $i, j$. This assumption does not take into account a possible network structure and thus useful statistical information may be lost. Network models such as the ones proposed in \cite{wang_ba2015} or \cite{billio_etal2019} fix $\underline{\pi}=2/(M-1)$ or estimate $\underline{\pi}$. 

\subsubsection{Combining SBMs and spike and slab priors}\label{sssec: SBM_SL}
In this paper we embed a network structure into an SSVS prior through individual inclusion probabilities $\underline{\pi}_{i,j}$. Since estimating $N = M (M-1)/2$ separate inclusion probabilities might lead to overfitting, we exploit the notion that shocks tend to cluster together. To achieve this,  we assume that a binary adjacency matrix $\bm \Delta$ with generic element $\delta_{i,j}$ arises from a stochastic block model \citep[SBM, see][]{holland_etal1983, nowicki_snijders2001}. Instead of assuming that the indicators $\delta_{i,j}$ are i.i.d. Bernoulli distributed with a common success probability $\underline{\pi}$, we allow for a latent network involving a SBM to govern the contemporaneous relationships among the shocks. 

We start our discussion by introducing terminology necessary to describe networks. We assume that the time-invariant network encoding  can be represented by a latent, random graph $\mathcal{G}$ which in our case encodes possible networks that give rise to $\bm \Omega$. We furthermore assume that $\mathcal{G}$ is undirected (i.e. $\delta_{i,j} = \delta_{j,i}$) and unweighted (all links are of the same magnitude). It is given by the duplet $(\mathcal{V,E})$, where $\mathcal{V} = 1,\dots,M$ denotes the vertex set (the members of the network) and $\mathcal{E} \subset \mathcal{V} \times \mathcal{V}$  the edge set (the connections among its members) which has cardinality $N$.

The graph is random in the sense that the edge set is generated by a Bernoulli process, but we assume the vertex set to be fixed. A binary representation of the graph can be obtained through the adjacency matrix $\bm \Delta$ which has elements $\delta_{i,j}=1$ if an edge exists between nodes $i$ and $j$ and $\delta_{i,j}=0$ otherwise. It is worth stressing that, in contrast to papers such as \cite{legramanati_etal2022}, our adjacency matrix is latent and controls whether the elements in $\bm \Omega$ are shrunk to zero or not.

In the case of an SBM $\mathcal{G}$ is driven by a community structure that clusters the nodes into $H$ distinct groups. For typical macroeconomic applications, these clusters could represent country groups (e.g., emerging and developed economics), variable types (e.g., prices, labor market quantities or financial series) or types of shocks (e.g., demand versus supply shocks; common versus idiosyncratic shocks). We do not assume any such known structure, but rather rather estimate both the membership of each group and the number of groups. 


Let $h,k \in \{1,\dots,H\}$ denote the group memberships of nodes $i$ and $j$, respectively. An edge exists between $i$ and $j$ with probability $\pi_{h,h} \in [0, 1]$ if $i$ and $j$ are members of the same group $h$. If the groups of $i$ and $j$ differ, an edge exists with probability $\pi_{h,k}$.  This implies that relations within as well as across groups are assumed to be homogeneous a priori. A typical assumption  is that $\pi_{h,h} \gg \pi_{h,k}$, such that vertices within the same group are more likely to be connected with one another. 

\begin{figure}[!tbp]
    \centering
    \begin{subfigure}{0.3\textwidth}
        \centering
        \includegraphics[width=\textwidth,scale=0.5]{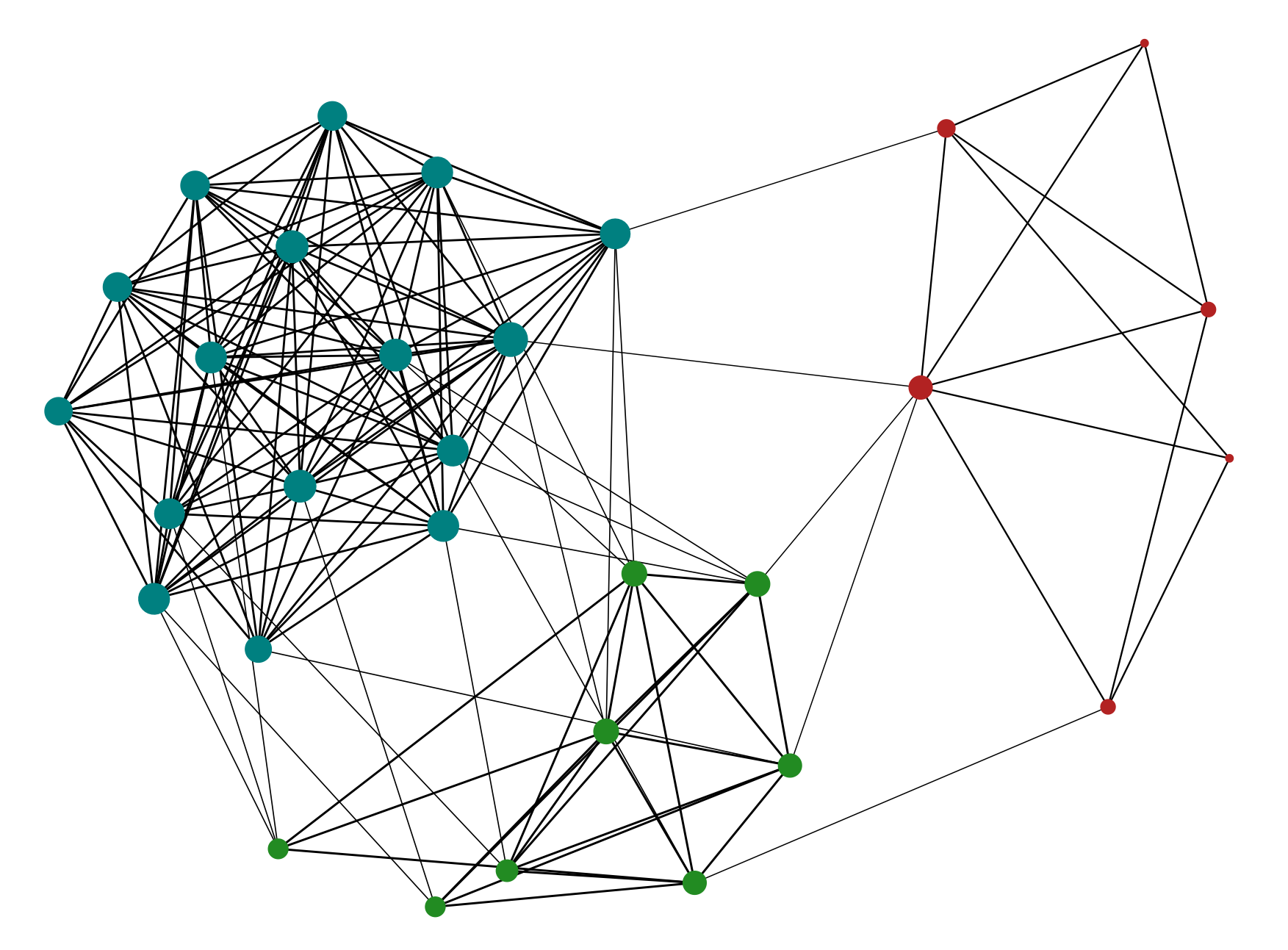}
        \caption{}
        \label{}
    \end{subfigure}
    \hfill
    \begin{subfigure}{0.3\textwidth}
        \centering
        \includegraphics[width=\textwidth,scale=0.5]{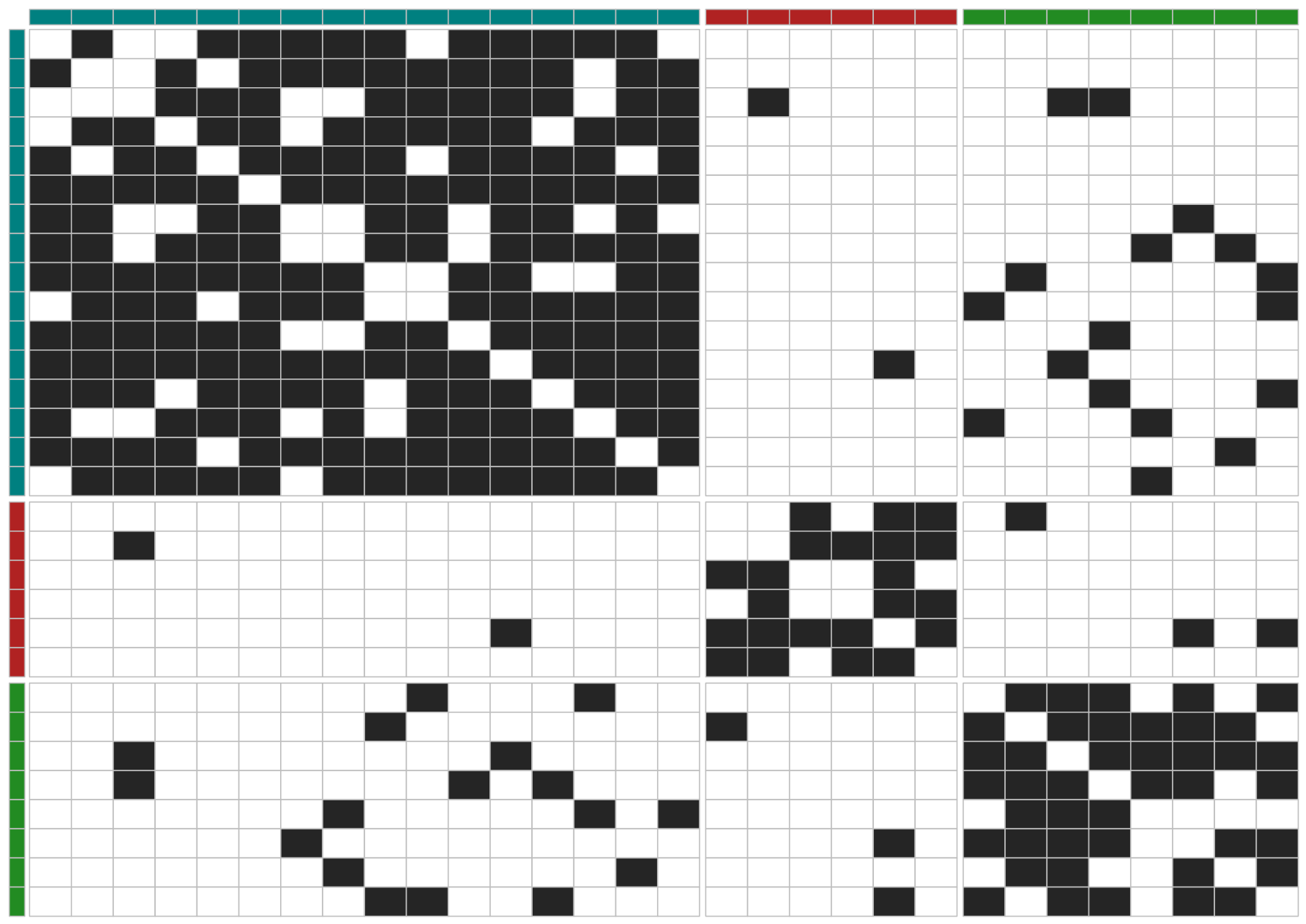}
        \caption{}
        \label{}
    \end{subfigure}
    \hfill
    \begin{subfigure}{0.3\textwidth}
        \centering
        \includegraphics[width=\textwidth,scale=0.5]{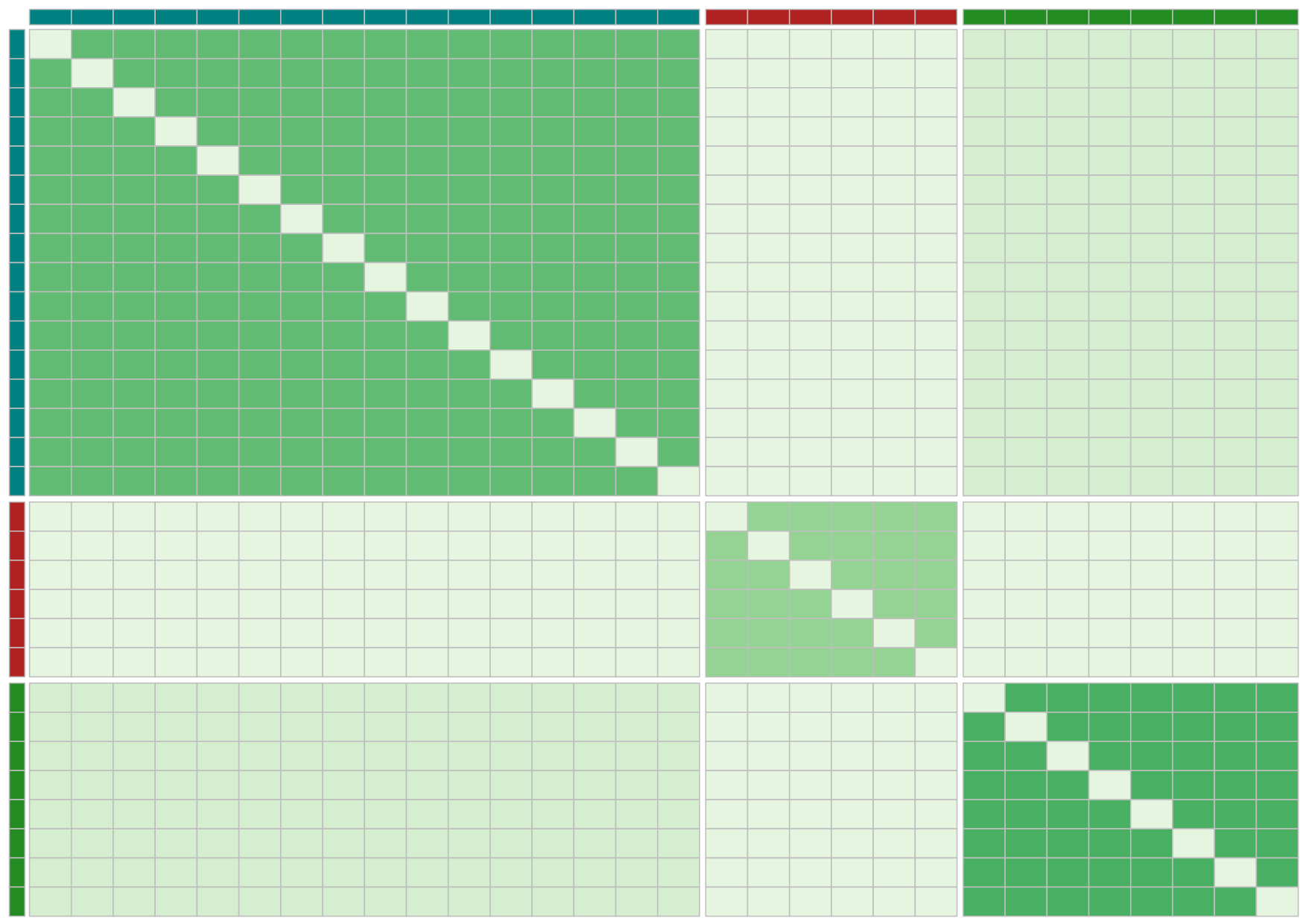}
        \caption{}
        \label{}
    \end{subfigure}
    \caption{Realization of a stochastic block model with $H=3$ and $M=30$. Node and annotation colors indicate group membership. (a) Plot of the network (thicker lines indicate higher edge probabilities). (b) Heatmap representation of the adjacency matrix. (c) Heatmap of the edge probabilities (darker cells indicate higher edge probabilities).}
    \label{fig:sample_nw}
\end{figure}

Figure \ref{fig:sample_nw} displays an example of such a network with $M=30$ nodes and $H=3$ groups alongside heatmaps of the implied adjacency matrix and edge probabilities. The key feature is that variables which display similar properties are grouped together in one of three groups.  These groups, however, arise endogenously and we will use this feature to design our prior for the elements of $\bm \Omega$.  Before discussing the mathematical structure of the model it is also worth emphasizing that the number of groups $H$ is not fixed a priori. In what follows we let the model decide on the number of groups and estimate it alongside the remaining model parameters.

Our approach builds on the Extended Stochastic Block Model (ESBM) framework put forward by \cite{legramanati_etal2022}. It extends the original SBM established by \cite{holland_etal1983} and \cite{nowicki_snijders2001} by providing a unified modeling framework and an efficient sampling algorithm for a range of different model formulations, with full uncertainty quantification and without the need to pre-specify the number of clusters $H$. 

The SBM implies a conditional prior on the latent adjacency matrix $\bm \Delta$:
\begin{equation*}
    p(\bm \Delta | \bm \Pi) = \sum_{h=1}^{H} \sum_{k=1}^h \pi_{h,k}^{m_{h,k}} (1-\pi_{h,k})^{\Bar{m}_{h,k}},
\end{equation*}
Here, $\bm \Pi$ is a $H \times H$ symmetric matrix with generic elements $\pi_{h,k}$. $m_{h,k}$ and $\Bar{m}_{h,k}$ denote the number of edges and non-edges between groups $h$ and $k$, respectively. 

Notice that this prior only depends on the edge probabilities across groups and not across nodes. This is the main source of shrinkage in our model and requires the introduction of a set of group assignment indicators  $\bm z = (z_1, z_2, \dots, z_M)$, which is an $M$-dimensional vector. Each element of $\bm z, z_j$, is a discrete random variable which takes values $1, \dots, H$.  The node-specific probabilities can then be obtained as $\underline{\pi}_{ij} = \pi_{z_i, z_j}$.

We assume that the random discrete vector $\bm z$ arises from a general Gibbs-type prior \citep{legramanati_etal2022}:\footnote{\cite{legramanati_etal2022} argue that  previous SBMs \citep[e.g.,][]{nowicki_snijders2001, kemp_etal2006, schmidt_morup2013, geng_etal2019} can be interpreted as arising from the use of different Gibbs-type priors.}
\begin{equation}
    p(\bm z) = \mathcal{W}_{M,H} \prod_{h=1}^H (1-\sigma)_{n_h-1},
\end{equation}
where $(a)_n$ for any $a>0$ denotes the ascending factorial $a (a+1) + \dots + (a+n-1)$, $n_h$ is the number of members in group $h$, $\sigma < 1$ a discount parameter and $\{\mathcal{W}_{M,H}: 1\leq H \leq M\}$ a collection of non-negative weights satisfying $\mathcal{W}_{M,H} = (M-H \sigma) \mathcal{W}_{M+1,H} + \mathcal{W}_{M+1,H+1}$ and $\mathcal{W}_{1,1}=1$. 
This general formulation  nests a broad range of priors on the clustering structure that can accommodate different types of networks.\footnote{See \cite{lijoi_etal2007a, lijoi_etal2007b, deblasi_etal2013, deblasi_etal2015, miller_harrison2018} for reviews and examples of various priors that fall into this category.}

We consider various Gibbs-type priors in this paper. In particular, we use the Gnedin-process \citep[GN;][]{gnedin2010}, Dirichlet-Multinomial \citep[DM;][]{nowicki_snijders2001}, Dirichlet-process \citep[DP;][]{kemp_etal2006} or Pitman-Yor-process \citep[DP;][]{pitman_yor1997_aop} priors.
Different choices of Gibbs-type priors lead to different grouping structures. As discussed in \cite{legramanati_etal2022}, the DM prior typically generates modular, relatively stable clusters with a fixed number $\Bar{H}$ of population clusters. The GN prior can be seen as a generalization thereof, where $\Bar{H}$ is modeled as random but still assumed to be finite in population. The DP and PY priors on the other hand favor more fragmented networks and smaller groups with no limit on $\Bar{H}$ as $M$ increases. They differ in the growth of $H$, which is much faster for the PY than for the DP prior. When specifying the hyperparameters for the Gibbs type priors, a sensible approach is to choose them such that their expected number of clusters matches the researcher's prior beliefs.

The SBM prior is completed by specifying a prior on $\pi_{h, k}$. We follow \cite{legramanati_etal2022} and specify independent Beta-distributed priors on $\pi_{h,k} \sim \textit{}{Beta}(a_\pi, b_\pi)$. Setting $a_\pi=b_\pi=1$ induces a uniform prior bounded by zero and one. 
 
\subsubsection{Priors on the remaining parameters}
In this sub-section, we briefly sketch the remaining priors on the model parameters. This includes the prior on the VAR coefficients and the prior on the parameters in the state equation of the log-volatilities.

The prior on  the VAR coefficients in $\bm a = \text{vec}(\bm A_1, \dots, \bm A_p)$ matrices in (\ref{eq:var}) is a standard Horseshoe shrinkage prior \citep[see, e.g.,][]{carvalho_etal2010}. The Horseshoe implies a hierarchical Gaussian prior on each element in $\bm a$:
\begin{equation*}
    a_j \sim \mathcal{N}(0, c_j^2 d^2), \quad c_j \sim \mathcal{C}^+(0, 1), \quad d \sim \mathcal{C}^+(0,1).
\end{equation*}
Here, we let $c_j$ denote a local shrinkage parameter and $d$ a global shrinkage factor. Both have half-Cauchy distributed prior distributions. We store these parameters in a $K (=M^2 P)+1$ vector $\bm \xi = (c_1^2, \dots, c_{M^2 P}^2, d)'$. We rely on this prior given that it works well in large dimensions and its recent success in many macroeconomic forecasting applications \citep{follet_yu2019, feldkircher_etal2022_ier, bai_etal2022_jae} but any other large VAR prior could be used without altering the main themes of this paper. 

On the parameters of the state equation of the log-volatilities we use a truncated Gaussian distribution with mean equal to $0.7$, variance of $0.1$ and support over $[-0.99,0.99]$ on the persistence parameter and a Gamma distributed prior  with shape parameter $10$ and rate parameter equal to $2$ on the error precision of the state equation.

\subsection{Posterior simulation}
In this sub-section we discuss how to carry out posterior simulation in the SBM-VAR. 

The joint posterior distribution of the  parameters and latent states is given by:
\begin{align*}
    p(\bm a, \bm \xi, \{\bm d_t\}_{t=1}^T, \bm \Omega, \bm \Delta, \{\tau_{i, j}\}_{i, j}, \bm \Pi, \bm z| \bm Y)
\end{align*}
where $\bm Y = (\bm y_1, \dots, \bm y_T)'$. This joint distribution is not available in closed form.  However, for most parameters we have conditional distributions that take a well known form and hence are amenable to Gibbs sampling. Here, we provide a detailed overview of the algorithm. Additional details  are provided in Appendix \ref{app:full_posterior}.

Our MCMC algorithm cycles between the following steps:
\begin{enumerate}
    \item \textbf{Sampling from }$p(\bm a|\bm \xi, \{\bm d_t\}_{t=1}^T, \bm \Omega, \bm Y)$. We sample the rows of $\bm A = (\bm A_1, \dots, \bm A_p)$ using the triangularization algorithm developed in  \cite{carriero_etal2022}. Each of the $M$ rows, labeled $\bm a_{j, \bullet}~(j=1,\dots,M)$ is normally distributed so that:
    \begin{equation*}
        \bm a_{j, \bullet}| \bm A_{-j, \bullet}, \bm \xi, \{\bm d_t\}_{t=1}^T, \bm \Omega, \bm Y \sim \mathcal{N}(\overline{ \bm a_{j, \bullet}}, \overline{\bm V}_{a, j}),
    \end{equation*}
    with $ \bm A_{-j, \bullet}$ being the matrix $\bm A$ with the $j^{th}$ row excluded and
    $\overline{ \bm a_{j, \bullet}}$ and $\overline{\bm V}_{a, j}$ denoting the posterior mean and variance, respectively. Precise forms of these can be found in Appendix \ref{app:full_posterior}. Repeating this step for each row provides  a valid draw from $p(\bm a|\bm \xi, \{\bm d_t\}_{t=1}^T, \bm \Omega, \bm Y)$.

    \item \textbf{Sampling from }$p(\bm \xi|\bm a, \bm Y)$. The prior variances $\bm \xi$ are simulated based on the algorithm proposed in \cite{makalic_schmidt2016}. This involves introducing additional auxiliary parameters which are then sampled from inverse Gamma conditionals. Based on these, the conditional posteriors of $\{c^2_j\}_{j=1}^{M^2P}$ and $d^2$ are inverse Gamma as well. More details can be found in Appendix \ref{app:full_posterior}.
    
    \item \textbf{Sampling from }$p(\{\bm d_t\}_{t=1}^T | \bm a, \bm \Omega, \bm Y)$. We update $\bm d_t$ for $t=1, \dots, T$  using  the single-move algorithm outlined in \cite{ishihara_omori2012_csda}. In our simulations and real data applications this simple sampler works  well, yielding satisfactory mixing properties.

    \item \textbf{Sampling from }$p(\bm \Omega | \bm a, \{ \bm d_t \}_{t=1}^T, \bm \Delta, \{\tau_{i, j}\}_{i,j}, \bm Y)$. Generating draws for $\bm \Omega$ is made difficult by the fact that it needs to be positive definite. \cite{wang2012_ba, wang_ba2015} put forward a change-of-variable approach to sequentially sample the rows and columns of $\bm \Omega$. Let $\bm S = \Tilde{\bm \varepsilon}' \Tilde{\bm \varepsilon}$ where $\Tilde{\bm \varepsilon} = (\bm D_1^{-1} \bm \varepsilon_1, \dots, \bm D_T^{-1} \bm \varepsilon_T)'$ are the VAR residuals rescaled using the time-varying part of the error volatility. Furthermore, let $[\underline{\bm V}_\Omega]_{ij}=\delta_{i,j} \tau_{i,j}^2 + (1-\delta_{i,j}) \tau_{ij,0}^2$ be the prior variance matrix for $\bm \Omega$. To sample column $j\ (j=1,\dots,M)$ we permute $\bm \Omega,\ \bm S$ and $\underline{\bm V}_{\Omega}$ such that $j$ is ordered last. We denote the permuted matrices $\Tilde{\bm \Omega},\ \Tilde{\bm S}$ and $\Tilde{\underline{\bm V}}_{\Omega}$ and apply the following decomposition:
    \[
      \Tilde{\bm \Omega} = \bigl(\begin{smallmatrix}
        \bm \Omega_{(-j)(-j)} & &  \bm \omega_{(-j)j} \\
        \bm \omega_{(-j)j}' & &  \omega_{jj}
      \end{smallmatrix}\bigr),\
      \Tilde{\bm S} = \bigl(\begin{smallmatrix}
        \bm S_{(-j)(-j)} & & \bm s_{(-j)j} \\
        \bm s_{(-j)j}' & &  s_{jj}
      \end{smallmatrix}\bigr),\ 
      \Tilde{\underline{\bm V}}_\Omega = \bigl(\begin{smallmatrix}
        \underline{\bm V}_{\Omega,(-j)(-j)} & & \underline{\textbf{\textsc{v}}}_{\Omega,(-j)j} \\
        \underline{\textbf{\textsc{v}}}_{\Omega,(-j)j}' & & \underline{\textsc{v}}_{\Omega,jj}
      \end{smallmatrix}\bigr),
    \]
    where $\bm \Omega_{(-j)(-j)}$ are all columns and rows of $\bm \Omega$ except for the $j$-th, $\bm \omega_{(-j)j}$ is the $j$-th column of $\bm \Omega$ without its $j$-th element and $\omega_{jj}$ is the $j$-th diagonal element of $\bm \Omega$.

    As shown in \citet{wang2012_ba, wang_ba2015} it is possible to introduce two auxiliary variables $\bm u$ and $v$ whose distributions have a well known form. In particular, we sample from the following conditional posterior distributions:
    \begin{align*}
        v &\sim \text{Ga}\Bigr(\frac{T}{2} +1, \frac{s_{jj}}{2}\Bigl), \\
        \bm u &\sim \mathcal{N}\Bigr(-\bm C \bm s_{(-j)j}, \bm C \Bigl),
    \end{align*}
    where the Gamma distribution is indicated by $\text{Ga}(\cdot,\cdot)$ and  $\bm C = (s_{jj} \bm \Omega_{(-j)(-j)}^{-1} + \text{diag}(\underline{\textbf{\textsc{v}}}_{\Omega,(-j)j})^{-1})^{-1}$.\footnote{A random variable $z$ follows a Gamma distribution,  with density $p(z|\alpha,\beta) = \frac{\beta^\alpha}{\Gamma(\alpha)} (z)^{\alpha-1} \text{exp}(-\beta z)$.} Having obtained a draw for $(\bm u, v)$, we map them into $(\bm \omega_{(-j)j},\omega_{jj})$ via: 
    \begin{align*}
        \bm \omega_{(-j)j} &= \bm u, \\
        \omega_{jj} &= v + \bm \omega_{(-j)j}'\bm \Omega_{(-j)(-j)}^{-1} \bm \omega_{(-j)j}.
    \end{align*}
    We perform these steps sequentially for $j=1,\dots,M$ to obtain an update for each column and row of $\bm \Omega$.
    
    \item \textbf{Sampling from } $p(\bm \Delta|\bm \Omega, \{\tau_{i,j}\}_{i, j}, \bm \Pi, \bm z)$. Equation (\ref{eq:ssvs}) can be used to infer $\delta_{ij}$ \citep{ahelegbey2016_aas, ahelegbey_etal2016_jae}. The resulting posterior distribution $p(\delta_{i,j}|\omega_{i,j},\underline{\pi}_{i,j})$ is a Bernoulli distribution and $\bm \Delta$ can be constructed by sampling from $p(\delta_{i,j}|\omega_{i,j},\underline{\pi}_{i,j}) \sim \textit{Bernoulli}(\overline{\pi}_{i,j})$ with
        \begin{equation*}
            \overline{\pi}_{i,j} = \frac{\mathcal{N}(\omega_{i,j}| 0, \tau_{i,j}^2) \cdot \underline{\pi}_{i,j}}{\mathcal{N}(\omega_{i,j}| 0, \tau_{i,j}^2) \cdot \underline{\pi}_{i,j} + \mathcal{N}(\omega_{i,j}| 0, (c \cdot \tau_{i,j})^2) \cdot (1-\underline{\pi}_{i,j})}.
        \end{equation*}
    
    \item \textbf{Sampling from } $p(\{\tau_{i,j}^2\}_{i, j}|\bm \Omega, \bm \Delta)$. We update these  by sampling from their inverse Gamma distributed conditional posterior distributions. More details are provided in Appendix A. 
    
    \item \textbf{Sampling from }$p(\bm \Pi, \bm z | \bm \Delta)$. We sample the group-specific edge probabilities in the group assignments in two steps. First, we sample the group assignments marginally of $\bm \Pi$ from:
    \begin{equation*}
        \bm z \sim p(\bm z | \bm \Delta).
    \end{equation*}
    This step is carried out using Algorithm 1 proposed in  \cite{legramanati_etal2022}. More information are provided in Appendix A. 
    
    Conditional on $\bm z$ we sample the group-specific probabilities from Beta distributions:
    \begin{equation*}
          \pi_{h, k} \sim \mathcal{B}(\pi_{h, k}|a_\pi+ m_{h,k}, b_\pi + \overline{m}_{h, k}), \quad \text{ for } h,k = 1, \dots, H,
    \end{equation*}
    where $h,k=1,\dots,H$ are group indicators and $m_{h,k} (\Bar{m}_{h,k})$ denotes the number of links (missing links) between clusters $h$ and $k$.
    
    Notice that this step implies that we first sample from the marginal (of $\bm \Pi)$ and then from the conditional (on $\bm z$). Hence, the ordering of the steps matter to obtain a valid draw from $p(\bm \Pi, \bm z | \bm \Delta)$.
\end{enumerate}
After a sufficiently long burn-in period, this algorithm yields draws from  $p(\bm a, \bm \xi, \{\bm d_t\}_{t=1}^T, \bm \Omega, \bm \Delta, \{\tau_{i, j}\}_{i, j}, \bm \Pi, \bm z| \bm Y)$.  In our simulation exercises and empirical work we repeat the sampler $15,000$ times and discard the first $5,000$ iterations as burn-in. Due to  space constraints and to further reduce the autocorrelation of the retained draws we keep only every other draw.

\section{Simulation-based evidence}
In this section we analyze how our approach performs in a controlled environment. We generate data using a VAR(1) model:
\begin{equation*}
   \bm y_t = \bm A \bm y_{t-1} + \bm \varepsilon_t, \quad \bm \varepsilon_t \sim \mathcal{N}(\bm 0_M, \bm \Sigma_t), \quad \bm \Sigma_t = \bm D_t \bm \Sigma \bm D_t, \quad t=1, \dots, T 
\end{equation*}
and $\bm y_0 = \bm 0_M$. We assume that each element of $d_{j, t}~(j=1, \dots, M)$ evolves according to an AR(1) process $d_{j, t} = 0.9 d_{j, t-1} + \sqrt{0.2} \nu_{j, t}$ with $\nu_{j, t} \sim \mathcal{N}(0, 1)$ and $d_{j,0}=0$. This VAR(1) DGP is then set up so as to differ
 along several dimensions. First, we analyze model performance for differently sized models. The different sizes are $M \in \{5,  30, 50\}$, covering small, medium and large-sized datasets. Second, we consider the role of the length of the sample $T$. Here, we set $T \in \{200, 300, 400, 500\}$ to reflect situations commonly encountered when working with quarterly, post-war US data (i.e. so that $T$ is between $200$ and $300$) as well as monthly series (with $T$ being $400$ or $500$). Third, the DGPs assume a network structure so that the true matrix $\bm \Omega$ under the DGP has a particular structure which the SBM can uncover. The true $\bm \Omega$ is generated in the following way: Each of the variables is assigned to a cluster (with the number of true clusters depending on $M$). We draw the within-group edge probabilities from the Beta distribution $\text{Beta}(1,100)$ and the cross-group edge probabilities from $\text{Beta}(100,1)$. The elements of the adjacency matrix are generated by drawing from a Bernoulli distribution with the respective edge probability between variables $i$ and $j$ as success probability. Finally, we  generate $\bm \Omega$ as a positive definite matrix with zero restrictions given by the adjacency matrix. We also consider the case of no network structure. In this case, the edge probabilities are set equal to $0.2$.

This gives us a combination of $24$ different DGPs. For each of these $24$ DGPs we generate $25$ datasets and  evaluate the performance in terms of hit rates (HRs), i.e., correctly classified edges in the network, of our BVAR-SBM against a benchmark model with a standard SSVS prior on the covariances (henceforth labeled SSVS). 

Our SBM-VARs differ in the clustering priors used. We consider each of the four Gibbs-type priors mentioned in Section \ref{sssec: SBM_SL}: the GN, the DM, the DP and the PY prior. Their hyper parameters are specified such that the prior expected number of groups is equal to $1.5$ times the true number of groups (rounded to whole numbers). 

\begin{figure}[!hbt]
    \centering
    \includegraphics[scale=0.8]{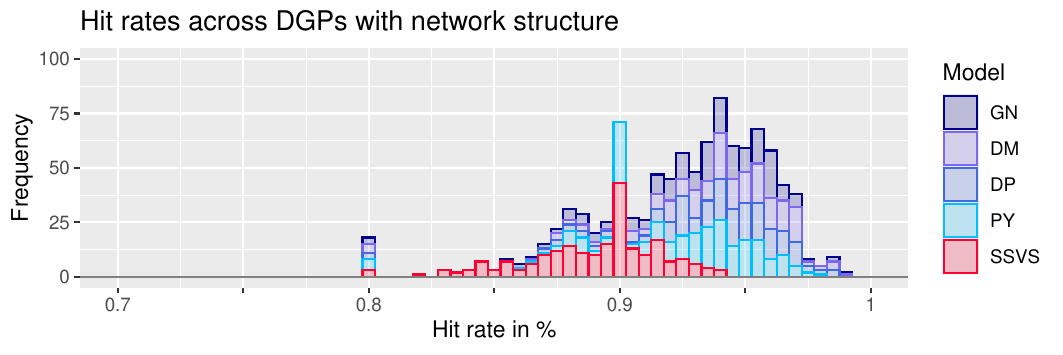}
    \caption{Percentage of correctly classified edges of different SBM-VARs  and the SSVS-VAR (red) over all draws from all DGPs with underlying network structure.}
    \label{fig:hist_hr}
\end{figure}

We start out by discussing the absolute HRs of the posterior median  $\overline{\bm \Delta}$ of $p(\bm \Delta|\bm Y)$ to the true value of $\bm \Delta$. Before we zoom into the DGP-specific performance, we consider the overall performance over different values of $T, M$ and across the realizations from the DGP. These are, in the form of histograms, shown in \autoref{fig:hist_hr} for the DGPs where $\bm \Omega$ features a network structure. In this figure, each of the histograms refers to one of the SBM priors (and the SSVS specification). The means of these histograms show the average quality of the network approximation (measured in HRs) over the space of DGPs.

\begin{figure}[!hbt]
    \centering
    \includegraphics[scale=0.8]{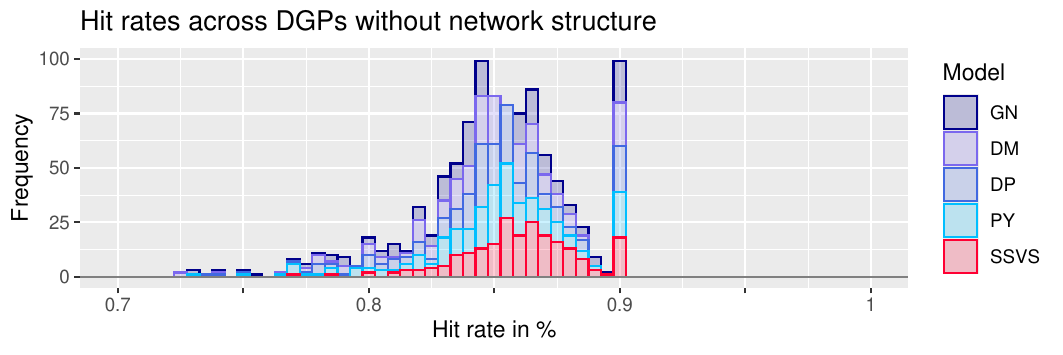}
    \caption{Percentage of correctly classified edges of different SBM-VARs  and the SSVS-VAR (red) over all draws from all DGPs with no underlying network structure.}
    \label{fig:hist_hr_nonw}
\end{figure}

The figure reveals that the different SBM priors all do well in terms of recovering the true network structure. For all models considered, the mean of the distribution of HRs is around 95 percent. By contrast, the SSVS prior that  assumes the prior inclusion probability as fixed across edges is performing significantly worse, with mean HRs that are appreciably lower than the ones of the SBMs. Notice that the left-tail of the distributions coincide, implying that all methods (including the SSVS) produce estimates of the network that have the same  lower bound on the HRs.

Next, we turn to the case where $\bm \Omega$ features no clustering. This is shown in Figure \ref{fig:hist_hr_nonw}. In this case, all SBM-based models produce hit rates that are close to the SSVS models. HRs are, however, markedly lower in all cases relative to the DGP that features clusters in $\bm \Omega$. Nevertheless, these results can be interpreted in a way that adding the SBM to the VAR does not harm network-detection accuracy if there is no particular clustering structure and, if there is clustering, it improves accuracy appreciably.

\begin{table}[h!]
\centering
    \begin{tabular}{cc c ccccc c ccccc}
        \toprule \toprule
        \textbf{M} & \textbf{T} & & \multicolumn{5}{c}{\textbf{Clustered DGPs}} & & \multicolumn{5}{c}{\textbf{Non-clustered DGPs}} \\ 
        & & & GN & DM & DP & PY & SSVS & & GN & DM & DP & PY & SSVS\\ \midrule
        5 & 200 & & 2.78 & 2.78 & 2.22 & 3.33 & 90.00 & & 0.00 & -1.00 & -2.00 & -1.00 & 93.50 \\ 
        30 & & & 4.00 & 4.14 & 3.93 & 3.94 & 88.74 & &  -0.60 & -1.02 & -0.76 & -0.69 & 86.73 \\ 
        50 &  & & 4.06 & 5.71 & 5.84 & 4.23 & 85.03 & &  -0.11 & -0.26 & -0.22 & -0.01 & 83.99 \\ \midrule
        5 & 300 & & 2.22 & 2.22 & 2.22 & 2.22 & 92.78 & &  -0.43 & -0.43 & 0.00 & -0.87 & 93.04 \\ 
        30 & &  & 3.69 & 4.04 & 3.95 & 3.61 & 90.24 & &  -1.31 & -1.66 & -1.61 & -1.10 & 86.84 \\
        50 & & & 5.47 & 6.83 & 6.13 & 5.69 & 87.15 & &  -0.50 & -0.70 & -0.63 & -0.38 & 85.92 \\\midrule
        5 & 400 &  & 0.62 & -0.62 & 0.00 & -0.62 & 94.38 & &  0.45 & 0.45 & 0.45 & 0.45 & 93.18 \\ 
        30 & &  & 2.79 & 2.99 & 2.98 & 2.81 & 91.74 & &  -3.01 & -2.47 & -2.81 & -2.72 & 85.43 \\ 
        50 & & & 5.72 & 6.79 & 6.45 & 5.62 & 88.56  & & -1.20 & -1.47 & -0.91 & -0.94 & 86.56 \\ \midrule
        5 & 500 &  & 0.53 & 0.53 & 0.53 & 0.53 & 93.68 & & -0.43 & 0.00 & 0.00 & 0.00 & 93.48 \\ 
        30 & & & 2.22 & 2.30 & 2.43 & 2.41 & 92.04 & &  -2.79 & -2.81 & -2.72 & -2.84 & 83.35 \\ 
        50 & &  & 5.61 & 6.75 & 6.04 & 5.29 & 89.64 & & -1.38 & -1.24 & -1.28 & -1.33 & 86.42 \\
        \bottomrule \bottomrule
    \end{tabular}
    \caption{Differences in hit rates between the different SBM-VARs and the SSVS-VAR. All results are means over 25 simulations from each DGP. For the SSVS-VAR absolute hit rates are reported.}
    \label{tab:tab_sim}
\end{table}

The histograms mask differences across different parameters of the DGPs. To understand under which circumstances our SBMs produce more favorable estimates of the network, we now zoom into the specific performance for each choice of $M, T$ and whether the DGP has a network structure or not. All the results are provided in Table \ref{tab:tab_sim}. The table shows the percentage point differences between the HR of a particular SBM-VAR and the SSVS-VAR.

We start our discussion by focusing on the absolute performance of the SSVS-VAR first. For all combinations of $T, M$ and for clustering and non-clustering DGPs, 
the SSVS-VAR produces hit rates well above 80 percent. For small models $(M=5)$, the model detects around 90 or more edges correctly. This is not surprising given the fact that the number of possible edges is $10$ and thus potential clusters are (at best) characterized by including relatively few nodes. When we increase the dimensionality of the model (i.e. set $M \ge 15$), the performance of the SSVS-VAR deteriorates and drops below 90 percent. Notice that using longer time series ($T \ge 300)$) improves network estimation accuracy across all model sizes.

The  SBM-VARs with different priors on the clustering behavior mostly improve upon the SSVS-VAR if the DGP features clustering. In most cases, differences across priors are small and within one to two percentage points.  When we consider the non-clustered DGP we find many values below zero. But these are small, suggesting that using the SBM does not hurt network detection accuracy if no clustering is present.

Zooming into the differences across priors for the non-clustered DGP we find that for some priors and DGPs, using an SBM produces (almost) the same HRs as in the case of the correctly-specified SSVS-VAR. 

To sum up, our simulations indicate that the choice of the prior on the clustering process is not critical. Nevertheless, we find that the DP is very often producing the highest HR for both DGPs.

\section{Macroeconomic forecasting using SBM-VARs}
The previous section shows that, with artificial data, our model is capable of recovering the underlying network of shocks accurately. Next, we ask whether explicitly modeling a network among the shocks also translates into more accurate density forecasts. We do so by using the different SBM-VARs to forecast US macroeconomic variables.

\subsection{Data and design of the forecasting exercise}
We rely on the quarterly version of the \cite{mccracken_ng2016_jbes} database from 1960:Q1 through 2023Q2. We consider $h=1$ and $h=4$ quarter ahead forecasts, calculated iteratively. Our forecast evaluation period starts in 1990:Q1. 

We consider two model sizes: a medium-sized dataset with $M=10$ and a larger one with $M=19$ endogenous variables. Both datasets include a set of three focus variables: these are the unemployment rate (UNRATE), the GDP growth rate (GDPC1) and CPI inflation (CPIAUCSL). Then, depending on the dataset we add additional macroeconomic and financial series. Further information on the actual series included in each model and how they are transformed are provided in the Data Appendix. 


To understand whether the prior on $\bm z$ matters for forecast accuracy, we again consider all four SBM priors and use the acronym SBM-VAR-X, where $\text{X} \in \{\text{GN}, \text{DM},\text{DP},\text{PY}\}$, to denote them. We also include the SSVS prior (without a network structure). All forecasting results are benchmarked against the Bayesian VAR without shrinkage on $\bm \Omega$ (i.e. a model that sets $\delta_{i,j} =1~\forall i,j$). This model is henceforth called the no shrinkage benchmark or baseline model. Thus we can separate out the gains from using some sort of SSVS prior shrinkage (by comparing SSVS results to the no shrinkage benchmark)  from the gains from using our specific SSVS network prior (by comparing the SBM-VAR-X models to SSVS). We stress that all these models share the same prior for the VAR coefficients and other model parameters and differ only in the prior for $\bm \Omega$.

Forecast accuracy is measured through log predictive likelihoods (LPLs). Since our goal is to model a network, we do not only focus on the univariate LPLs of the three focus variables but on joint LPLs.  These joint LPLs either measure the joint forecast density performance for the three focus variables (FOCUS) or for all variables in $\bm y_t$ (ALL).  

Before we discuss the empirical results, a brief word on how we set up the priors for $\bm z$. Note that the FRED-QD data divides the variables into groups (e.g. employment and unemployment, housing, etc.) and our data sets choose variables from some groups, but not others. The hyperparameters on the Gibbs priors are set so that the prior expectation on the number of groups equals the number of non-empty groups (for a particular model size). For instance, if the data set only includes labor market indicators and interest rates, we would tune the prior to expect only two clusters.

\subsection{Forecasting results}
\subsubsection{Overall density forecasting performance}
We start by discussing the overall forecast performance of the various VARs first. Table \ref{tab:lpls_1q_joint} shows the average LPLs of a particular model minus the one of the VAR without shrinkage on $\bm \Omega$. Positive numbers suggest that our model does better than the benchmark while negative numbers indicate the opposite. The column associated with 'BASE' shows the absolute LPL for the baseline specification. 

\begin{table}[!htbp] 
    \centering
    \resizebox{\columnwidth}{!}{%
    \begin{tabular}{c c cccccc c cccccc} 
        \\[-1.8ex]
        \toprule \toprule
        \textbf{Variable} & & \multicolumn{6}{c}{\textbf{Medium VAR}} & &  \multicolumn{6}{c}{\textbf{Large VAR}} \\
        \textbf{Group} & &  GN & DM & DP & PY & SSVS & BASE & & GN & DM & DP & PY & SSVS & BASE \\
        \midrule \\
        & \multicolumn{14}{c}{\textbf{1-quarter ahead}} \\[1.4ex]
        GDPC1 &  & 0.11 & 0.12 & 0.10 & 0.11 & 0.12 & 3.00 &  & 0.06 & 0.01 & 0.00 & 0.03 & 0.00 & 2.98 \\
        UNRATE &  & 0.24 & 0.12 & -0.08 & -0.20 & -0.07 & -7.62 & & 0.73 & -0.56 & -0.68 & -0.08 & -1.57 & -7.36 \\
        CPIAUCSL &  & 0.11 & 0.11 & 0.14 & 0.08 & 0.11 & 3.78 &  & 0.03 & 0.08 & 0.08 & 0.08 & 0.08 & 3.78 \\ \cmidrule{3-8} \cmidrule{10-15}
        FOCUS & & 0.73 & 0.45 & 0.29 & 0.16 & 0.35 & -0.87 & & 1.19 & -0.12 & -0.34 & 0.40 & -1.22 & -0.51 \\ 
        ALL & & 0.99 & 0.73 & 0.64 & 0.47 & 0.51 & & & -0.29 & -3.51 & -0.93 & -4.12 & -4.52 & \\
        \midrule \\
        & \multicolumn{14}{c}{\textbf{4-quarters ahead}} \\[1.4ex]
        GDPC1 &  & 0.04 & 0.01 & 0.04 & 0.06 & 0.03 & 2.50 & & 0.00 & -0.01 & 0.01 & 0.00 & 0.00 & 2.57 \\ 
        UNRATE &  & 0.11 & 0.29 & 0.10 & 0.15 & -0.01 & -6.15 & & 0.53 & 0.21 & 0.44 & 0.44 & 1.07 & -6.52 \\
        CPIAUCSL &  & 0.12 & 0.09 & 0.11 & 0.11 & 0.08 & 3.54 &  & 0.16 & 0.10 & 0.14 & 0.13 & 0.14 & 3.49 \\ \cmidrule{3-8} \cmidrule{10-15}
        FOCUS &  & 0.37 & 0.52 & 0.34 & 0.38 & 0.18 & 0.92 & & 0.85 & 0.52 & 0.80 & 0.74 & 1.52 & 0.36 \\ 
        ALL &  & 0.58 & 0.63 & 0.63 & 0.63 & 0.38 & &  & 2.42 & 2.45 & 3.53 & 2.81 & 1.91 & \\ 
        \bottomrule \bottomrule
    \end{tabular}
    }
    \caption{Joint LPL scores across different variable groups of the medium-sized and large VARs relative to the baseline model with no shrinkage. Scores of the baseline are in absolute terms.}
    \label{tab:lpls_1q_joint}
\end{table}

We start by discussing the one-step-ahead marginal LPLs of the medium-scale VAR first. For  two out of three focus variables (GDPC1 and CPIAUCSL), we find that SBM-VARs  and the SSVS-VAR improve upon the no shrinkage benchmark. In these cases, differences between the SSVS-VAR (which assumes no network structure in the prior inclusion probabilities) and the SBMs are muted. In particular, we find that for GDP growth, all models produce a similar performance, with the SBM-VAR-DM and the SSVS-VAR  producing identical LPL differences. For inflation, a similar pattern arises but here we find that the SBM-VAR-DP produces slightly more precise density forecasts than all competing models. Only for the unemployment rate we find that SBM-VAR-GN produces substantially more precise density forecasts than all other specifications. The unemployment rate is also the only focus variable where some of the SBM-VARs (and the SSVS-VAR) perform worse than the no shrinkage benchmark.

For the large model, a similar story arises. We find small gains for output and inflation. These, however, appear to be more pronounced when compared to the SSVS-VAR. For unemployment,  we find that the only model that improves upon the no shrinkage VAR is the SBM-VAR-GN with all other specifications performing worse than the benchmark. Comparing the marginal LPLs for the medium and large baseline model shows almost no differences. 

Intuitively, if our focus is on the marginal LPL for an individual variable, then the only way the SBM-VARs can show performance gains relative to SSVS is if they are better at modeling the contemporaneous spillovers between variables. Thus, it is unsurprising there are only small differences in marginal LPLs.  Joint LPLs should profit even more given that they depend on the full predictive covariance matrix $\bm \Sigma_{t+h}$.  

To back this claim, we start by discussing the joint LPL over the three focus variables. For the medium model, we find gains vis-\'{a}-vis the benchmark that are more sizable than the ones for any of the three focus variables. Again, the best performing specification is the SBM-VAR-GN with SBM-VAR-DM being second. In most cases (except for SBM-VAR-PY) we find that using the SBM to model $\bm \Omega$ pays off if the target is the joint density of the three focus series. When we consider the larger models we find a somewhat different story.  In this case, two out of four SBMs lose against the benchmark. But we also find that SBM-VAR-GN, again, substantially improves upon the benchmark and all other specifications (in particular the SSVS-VAR). This has to be considered in light of the absolute LPLs for the focus variables of the baseline model. Since these are slightly higher than the ones of the medium-scale VAR without shrinkage, we can say that if the researcher wishes to produce the most precise density forecasts for all three focus variables jointly, the single best performing specification is SBM-VAR-GN. 

Next, we consider the joint performance for all series in $\bm y_t$. If all series are considered, a similar pattern to the one based on the joint LPLs over the focus variables arises. With the medium-sized data set,  we find gains with respect to the benchmark with SBM-VAR-GN producing the most precise joint density predictions. For the large data set, interestingly, none of the models that induce shrinkage on $\bm \Omega$ manages to outperform the no shrinkage benchmark. In this case, the SBM-VAR-GN produces similar density predictions (which are, nevertheless, slightly worse than the ones of the benchmark).

Focusing on the one-year-ahead forecast distributions, a similar picture to the one-step-ahead horizon emerges. For marginals, the gains are rather small for output and inflation whereas they appear to be more substantial for the unemployment rate. This pattern holds for both model sizes. When we consider the LPLs over the focus series we again find SBM-VAR-GN to produce a strong performance, improving upon the benchmark and the SSVS-VAR. Notice, however, that the other priors perform slightly better. But these differences are muted. This also holds for the joint LPLs and when the medium-scale dataset is considered. 

For the large dataset, however, we find a different picture to the one observed for the one-quarter-ahead forecasts. In this case, we find that SBM-based models yield LPLs that are appreciably larger than the ones of the benchmark and the SBM-VAR-DP produces the most accurate density forecasts. 

\subsubsection{Density forecast performance over time}
At this point, we have discussed only average forecast performance over all periods in the hold out. To better understand why particular models are doing well we turn to analyzing the forecasting performance over time. 

\begin{figure}[t!]
\begin{minipage}[t]{0.48\textwidth}
    \centering \textbf{Medium}
\end{minipage}
\begin{minipage}[t]{0.48\textwidth}
    \centering \textbf{Large}
\end{minipage}\\
\begin{minipage}[t]{1\textwidth}
    \centering \textbf{ (a)  Focus variables}
    \vspace{.25cm}
\end{minipage}\\
    \centering
    \begin{subfigure}{0.49\textwidth}
        \centering
        \includegraphics[scale=0.38]{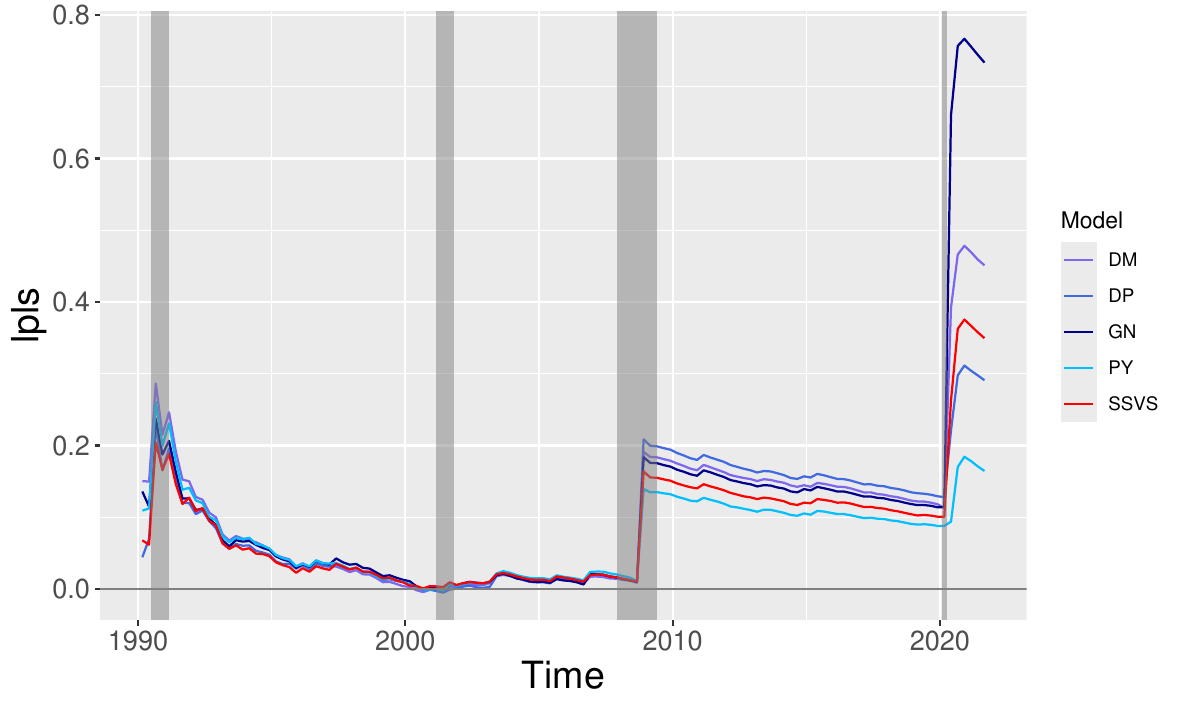}
        \label{subfig:lpls_nipa_small}
    \end{subfigure}
    \hfill
    \begin{subfigure}{0.49\textwidth}
        \centering
        \includegraphics[scale=0.38]{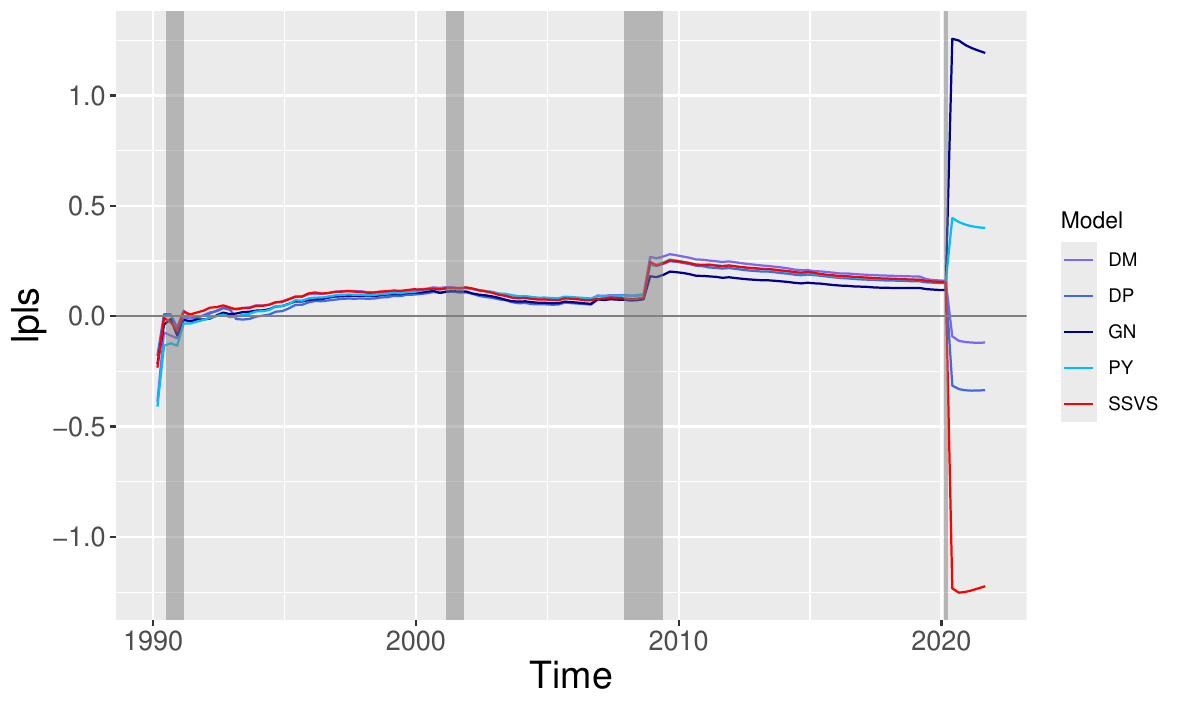}
        \label{subfig:lpls_empl_small}
    \end{subfigure}
    \begin{minipage}[t]{1\textwidth}
    \centering \textbf{ (b)  All variables}
    \vspace{.25cm}
\end{minipage}\\
    \begin{subfigure}{0.49\textwidth}
        \centering
        \includegraphics[scale=0.38]{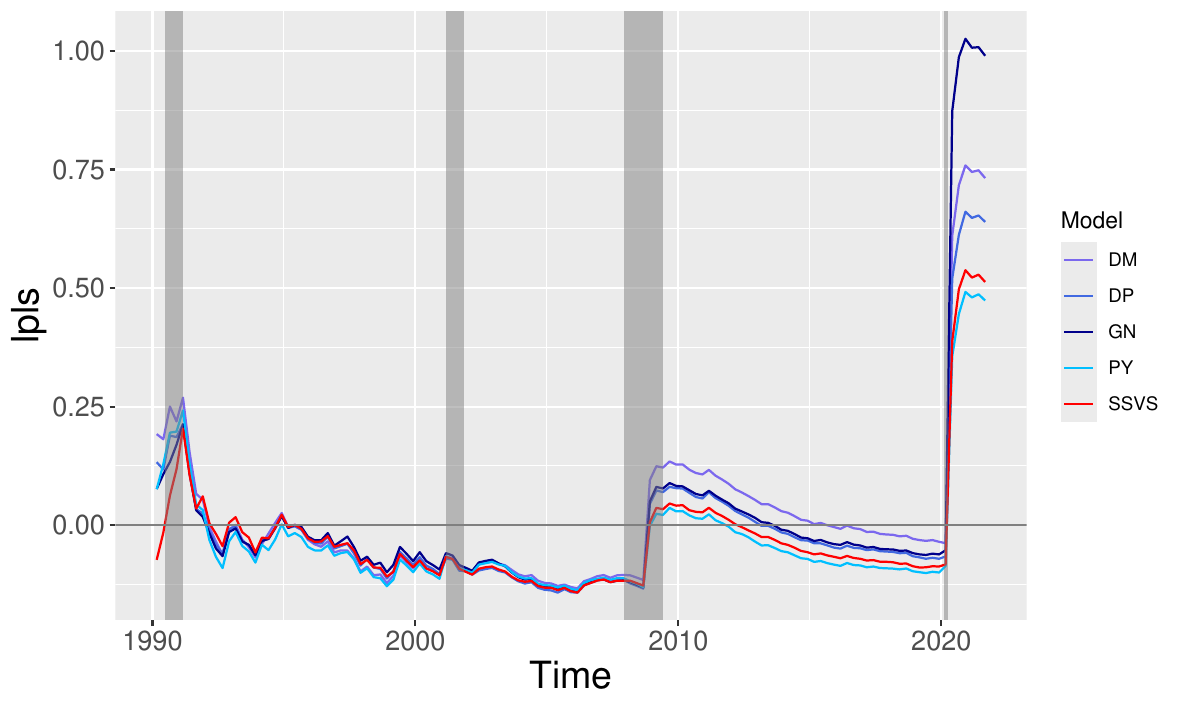}
        \label{subfig:lpls_focus_medium}
    \end{subfigure}
    \hfill
    \begin{subfigure}{0.49\textwidth}
        \centering
        \includegraphics[,scale=0.38]{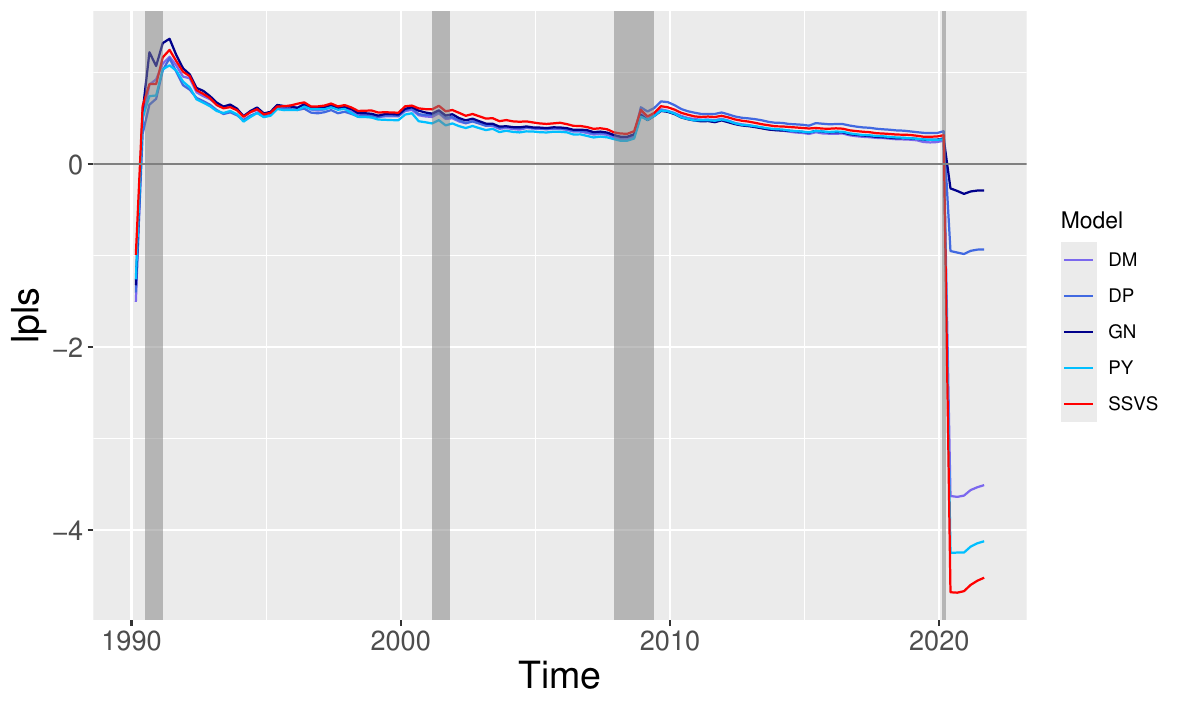}
        \label{subfig:lpls_joint_medium}
    \end{subfigure}
    \caption{Recursive mean of joint differences in one-quarter ahead LPLS scores relative to a VAR with a diffuse prior on the precision matrix. Shaded areas denote NBER reference recessions.}
    \label{fig:recursiveLPL_small}
\end{figure}

We do so in Figure \ref{fig:recursiveLPL_small}. This figure shows differences in recursive means between the different VARs that induce shrinkage on $\bm \Omega$ and the no shrinkage benchmark. At each point in time, a number greater than zero implies that a particular model has been outperforming the benchmark up to this point.  For brevity, we focus on the joint density over the focus variables and the joint density over all elements in $\bm y_t$ only.

Figure \ref{fig:recursiveLPL_small} strikingly shows that using models that induce shrinkage on $\bm \Omega$ helps forecast performance during volatile periods such as the global financial crisis or the pandemic. During both recessions, our models improve appreciably over the benchmark specification. In tranquil periods, this effect is much less pronounced and we even find that after the financial crisis, the relative differences somewhat decline up until 2020:Q1. 

This general pattern holds for both datasets and if we consider LPLs over the focus series and over all variables in the system. Notice, however, that the evolution of the relative LPLs is very similar for panels (a) and (b) for the medium data set, but differences are much larger if we compare panels (a) and (b) for the large dataset. In this case, two of the four SBM models (GN, DP) and the SSVS-VAR perform poorly during the pandemic (see panel (a)) but all of them fail to improve upon the benchmark if we consider joint LPLs over all variables in the system.

This discussion (and the discussion in the previous section) shows that our SBM-VARs are capable of improving upon the model that induces no shrinkage on $\bm \Omega$ and the VAR which assumes that the prior inclusion probabilities are fixed and independent from each other. Within the class of SBM-VARs, we find that SBM-VAR-GN does best for the one-step-ahead horizon while SBM-VAR-DP works best for four-quarter-ahead forecasts. In most cases, however, differences across the priors on $\bm z$ are rather small. 

\subsection{Inspecting network properties}
In this section, we investigate the properties of the estimated networks of the SBM-VARs with the GN and DP priors and the SSVS-VAR. Our focus on SBM-BVAR-GN and SBM-BVAR-DP is predicated on their strong performance for one and four-step-ahead predictive likelihoods, respectively. We analyze the networks by considering a set of summary statistics related to the network structure. These are computed recursively over the hold-out period.  For brevity, we focus on the medium-scale model.

We consider three different summary statistics:
\begin{enumerate}
    \item \textit{Number of groups.} This measures the number of groups $H$ detected by the SBM over the hold-out period. As a point estimate of $H$, we use the clustering which minimizes the posterior expectation of the variance of information loss function. \cite{wade_ghahramani2018_ba} discuss the advantageous properties of this approach.  
    \item \textit{Average degree.} The degree for each node $i$ is the number of edges between $i$ and all other nodes:
    \begin{equation*}
        d_i = \sum_{j\neq i} \delta_{i,j}.
    \end{equation*}
    The average degree is given by: $\Bar{d}=1/M\sum_{i=1}^M d_i$. The average degree measures the average number of links of a particular node to the other nodes in the network.

    \item \textit{Modularity.}  Modularity is a measure of how strongly separated from one another the different clusters are. It is defined as:
    \begin{equation*}
        Q = \frac{1}{2N} \sum_{i,j}\left(\delta_{i,j} - \frac{d_i d_j}{N} \right) \mathds{1}(z_i = z_j),
    \end{equation*}
    where $N$ is the overall number of edges and $\mathds{1}(z_i = z_j)$ is the indicator function that takes on value $1$ if nodes $i$ and $j$ are in the same groups, that is if $z_i = z_j$. If $Q$ is large, most edges in the network are within a particular cluster whereas if $Q$ is low, the community structure is weak and the edges are distributed uniformly over the network.
\end{enumerate}

\begin{figure}[!tb]
    \centering
    \begin{subfigure}{.49\textwidth}
        \centering
        \includegraphics[scale=.4]{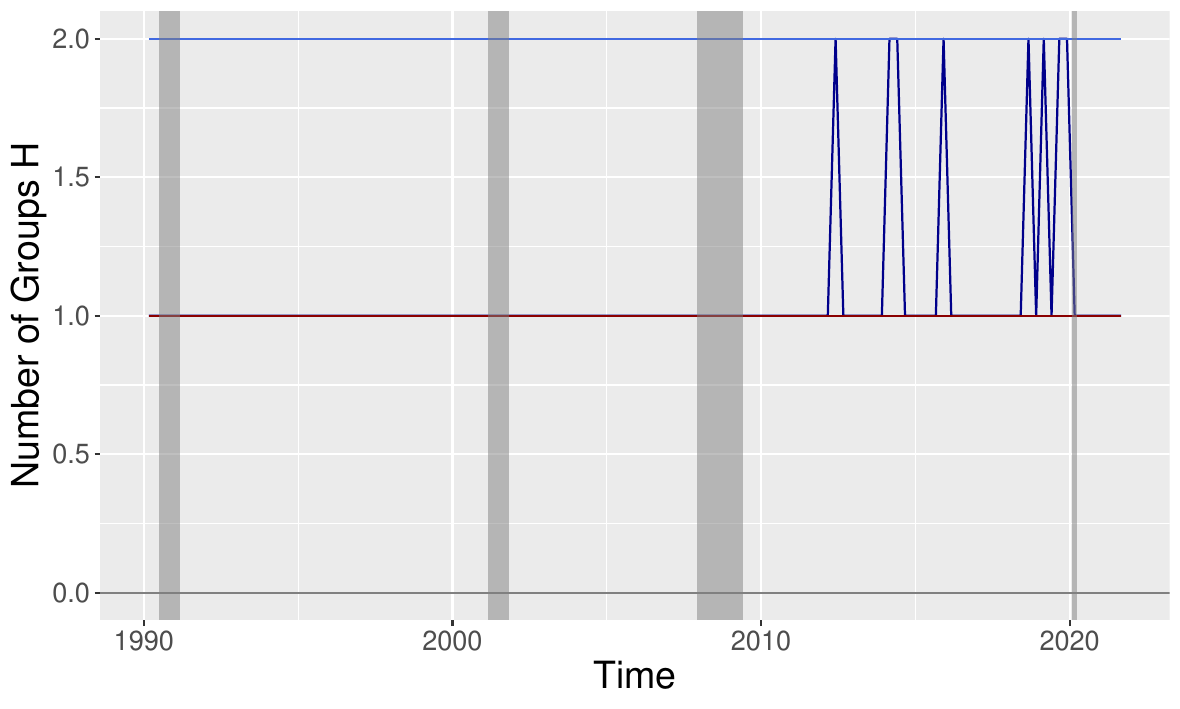}
        \caption{Number of groups}
        \label{subfig:no_groups1}
    \end{subfigure}
    \hfill
    \begin{subfigure}{.49\textwidth}
        \centering
        \includegraphics[scale=.4]{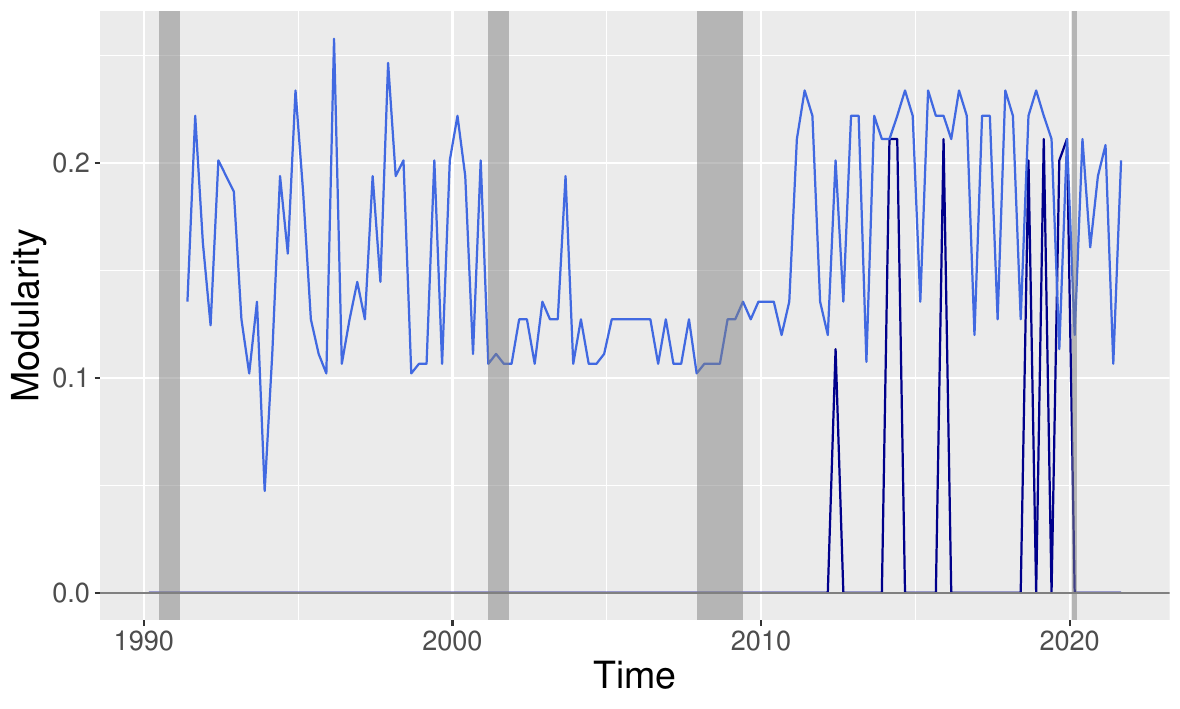}
        \caption{Modularity}
        \label{subfig:modul1}
    \end{subfigure}
    \vfill
    \begin{subfigure}{.49\textwidth}
        \centering
        \includegraphics[scale=.4]{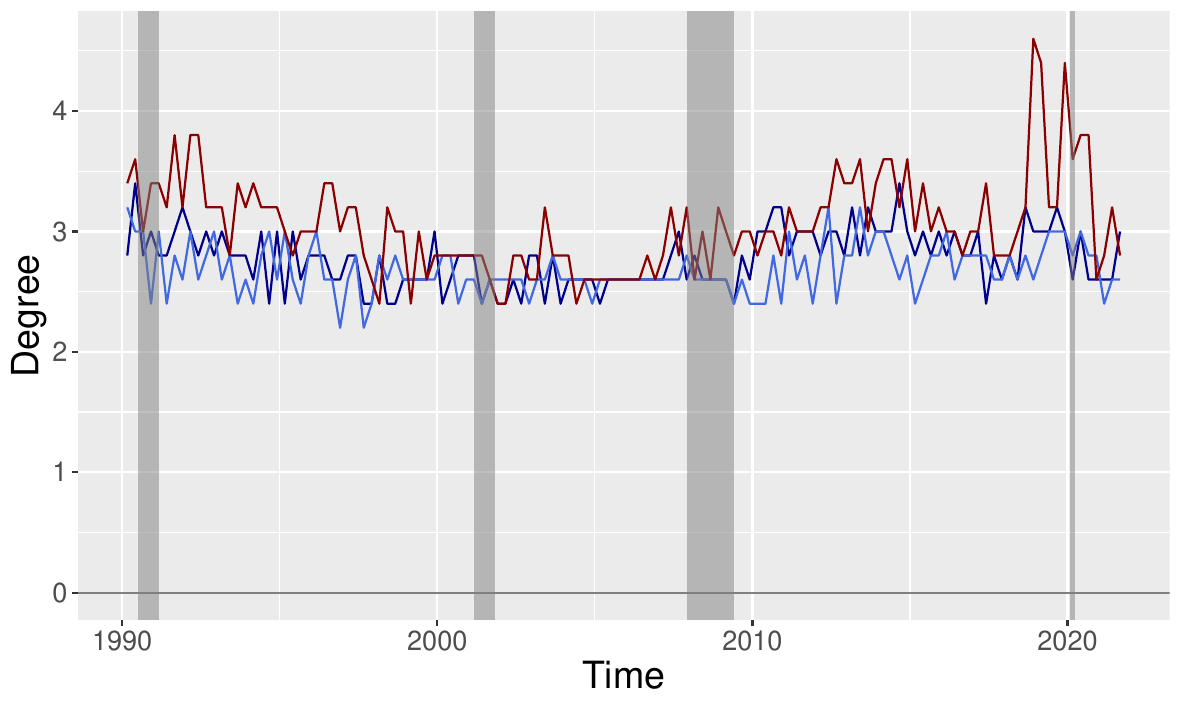}
        \caption{Average degree}
        \label{subfig:degree1}
    \end{subfigure}
    \caption{Network summary statistics over the holdout period. Dark blue is the SBM-GN prior, light blue the SBM-DP and red the SSVS prior. Shaded areas denote NBER reference recessions.}
    \label{fig:nw_summ}
\end{figure}

Figure \ref{fig:nw_summ} shows how these network summary metrics evolve over the hold-out period.  We start by discussing how the number of groups evolves over time.  Panel \ref{subfig:no_groups1} suggests that the model with the GN prior produces only a single cluster up until around 2012. Afterwards, the number of clusters fluctuates between one and two.  By contrast, when we consider the DP prior, which favors more dispersed partitions and tends to open up new clusters more quickly than the GN prior, the figure suggests two clusters throughout. The first cluster is comprised of price series and short-term interest rates whereas  the second cluster consists of the remaining series in the system.

Next, we consider network modularity, displayed in panel \ref{subfig:modul1}. Again, considering the GN prior first shows that modularity increases in lockstep with the number of groups.  It is noteworthy that modularity is appreciably lower in 2012 compared to the instances where the number of groups reach two in the later part of the hold-out period. This indicates that the structure of the groups has changed in the sense that community structure has been weaker in 2012 so that shocks within a group had a tendency to form more ties to shocks in the other group whereas for the cases where find two regimes later in the sample the connections have been strong within a particular group but less so between the two groups. Under the DP prior, we find that during the 1990s up until the burst of the dotcom bubble modularity is higher and decreases in the period between the dotcom crisis and the global financial crisis. During and after the financial crisis we find that cross-group relations become stronger.

Finally, we consider the average degree over time. This is shown in panel \ref{subfig:degree1}. We find that the network implied under the standard SSVS prior tends to be more dense than under the two SBM-type priors. For all three models, we observe that the degree is higher in the first third of the hold-out, before declining slightly during the early 2000s up to the financial crisis. In the aftermath of the financial crisis up to the pandemic the implied networks across all three priors tend to become more dense. This is particularly visible for the SSVS-VAR which gives rise to a rather dense $\bm \Omega$ matrix. 

\section{Conclusions}
Prior information can be important to ensure shrinkage and parsimony in large VARs. A substantial Bayesian VAR literature has developed proposing various  priors, but few of them focus on the high-dimensional error covariance matrix. In this paper we develop such a prior based on the idea that the error precision matrix is likely to exhibit a parsimonious network structure. It does so by using an SSVS prior (to select the parsimonious specification) with a SBM (which ensures the specifications involve networks).   We also develop a computationally efficient Bayesian MCMC algorithm which jointly estimates the number of clusters and the members in each cluster. 

In simulated data we find that this algorithm does a fine job of uncovering the network structure whereas a standard SSVS prior, designed to ensure parsimony but not involving a network structure, does not perform as well. This ability to find a parsimonious network in a high dimensional error precision matrix is also found in our forecasting exercise involving US macroeconomic data. In particular, we find a small number of clusters with sensible interpretations (e.g. grouping price and interest rate variables in one cluster and the remaining variables reflecting the real economy in another) but noticeable variation over time both in the membership of each cluster and in the strength of the links between members. Our forecasting exercise shows that the use of our SBM prior typically leads to some modest improvements over both an SSVS prior which ignores the network structure and a non-informative prior. 

As possible avenues for further research we stress that our model is very general and can be applied in combination with any specification for the conditional mean. Moreover,  the information we back out on the network can be used to inform shrinkage priors on the lagged coefficients as well. 

From an applied perspective, we illustrate our approach using US macroeconomic data. But in principle, we can use our framework to model financial time series such as stock returns which display a more pronounced clustering or herding behavior \citep[see, e.g.][for a multivariate SV model  that dynamically clusters series so that they exhibit herding/non-herding behavior]{tsionas2022multivariate}.

 \newpage

\bibliography{bib}

\newpage
\appendix
\section{Full Conditional Posterior Distributions}\label{app:full_posterior}

\noindent
\textbf{Sampling from} $p(\bm a|\bm \xi, \{\bm d_t\}_{t=1}^T, \bm \Omega, \bm Y)$. We let $\bm x_t = (\bm y_{t-1}, \dots, \bm y_{t-P})$ and stack the lagged observations over time to arrive at the $T\times (M~P)$-dimensional matrix $\bm X = (\bm x_1, \dots, \bm x_T)'$. Furthermore, $\bm B_0$ is the lower-Cholesky factor of $\bm \Omega$ and $\Tilde{\bm D}=(\bm d_1, \dots, \bm d_T)'$ a $T\times M$-matrix containing the sequence of time-varying observation error volatilities. \cite{carriero_etal2022} show that conditional on knowing $\bm B_0$ and $\bm \Tilde{\bm D}$ it is possible to factorize the model such that the VAR coefficients can be estimated equation-by-equation. In particular, for the $j^{th}$ equation ($j=1,\dots,M$) let:
\begin{align*}
    \bm Y^{(j)} &= \text{vec}((\bm Y-\bm X \bm A^{[j=0]})\bm B_{(0,j:M,\bullet)}')./\text{vec}(\Tilde {\bm D}_{\bullet,j:M}^{0.5}), \\
    \bm X^{(j)} & = (\bm B_{(0,j:M,j)} \otimes \bm X)./\text{vec}(\Tilde{\bm D}_{\bullet,j:M}^{0.5}),
\end{align*}
where $\bm A^{[j=0]}$ refers to a modified version of $\bm A$ where the $j$-th column is set equal to zero. The sub-matrix containing the rows with indices $j$ trough $M$ of $\bm B_0$ is denoted as $\bm B_{0,j:M,\bullet}$ and the vector $\bm B_{0,j:M,j}$ contains the elements $j$ to $M$ of the $j$-th column of $B_0$. Similarly, $\Tilde{\bm D}^{0.5}_{\bullet,j:M}$ are the columns $j$ through $M$ of $\Tilde{\bm D}^{0.5}$. The symbol $./$ indicates element-by-element division. Rewriting the variables in such a way, the moments of the conditional posterior arise from the standard Gaussian regression model:
\begin{align*}
    \overline{\bm V}_{a,j} &= (\underline{\bm V}_{a,j}^{-1} + \bm X^{(j)'} \bm X^{(j)})^{-1}, \\
    \overline{\bm a}_{j,\bullet} &= \Bar{\bm V}_{a,j}(\underline{\bm V}_{a,j}^{-1}\ \underline{\bm a}_{j,\bullet} + \bm X^{(j)'} \bm Y^{(j)}),
\end{align*}
where $\underline{\bm V}_{a,j}$ denotes the prior variance (determined by the elements of $\bm \xi$) and $\underline{\bm a_{j,\bullet}}$ the prior mean. For the simulations and empirical application we use a prior centered around zero.

\noindent
\textbf{Sampling from} $p(\bm \xi | \bm a)$. Recall that $\bm a = \text{vec}(\bm A_1, \dots, \bm A_p)$. The prior variance of each generic element $a_j\ (j=1,\dots,M^2P)$, is given by the product of a local, variable-specific component $c_j^2$ and a global, shared component $d^2$. \cite{makalic_schmidt2016} develop a simple sampling scheme for the case of the horseshoe prior involving two auxiliary variabels $\nu_j$ and $\zeta$. The conditional posterior of $c_j^2$ and $d^2$ takes the form:
\begin{align*}
    c_j^2 |\bullet &\sim \text{InvGa} \Bigl(1, \frac{1}{\nu_j} + \frac{a_j^2}{2 d^2}\Bigr), \\
    d^2 |\bullet &\sim \text{InvGa} \Bigl(\frac{M^2 P+1}{2}, \frac{1}{\zeta} + \frac{1}{2} \sum_{j=1}^{M^2 P} \frac{a_j^2}{c_j^2}\Bigr).
\end{align*}
The auxiliary variables are in turn sampled from:
\begin{align*}
    \nu_j | \bullet & \sim \text{InvGa} \Bigl(1, 1 + \frac{1}{c_j^2}\Bigr), \\
    \zeta | \bullet & \sim \text{InvGa} \Bigl(1, 1 + \frac{1}{d^2}\Bigr).
\end{align*}

\noindent
\textbf{Sampling from} $p(\{\tau_{i,j}^2\}_{i, j}| \{\bm d_t\}_{t=1}^T), \bm \Omega, \bm \Delta)$. Following \cite{ishwaran_rao2005} we update the prior variances of the off-diagonal elements of the symmetric precision matrix $\tau_{ij}^2 = \tau_{ji}^2\ (i=1,\dots,M; j = 1,\dots,i-1)$ by drawing from an inverse gamma distribution. The conditional posterior is given by:
\begin{equation*}
    \tau_{i,j}^2 \sim \mathcal{G}^{-1}\Bigr(a_\tau+\frac{1}{2},b_\tau + \frac{\omega_{i,j}^2}{\delta_{i,j} + (1-\delta_{i,j})c}\Bigl).
\end{equation*}

\noindent
\textbf{Sampling from }$p(\bm \Pi, \bm z | \bm \Delta)$. Conditional on the adjacency matrix $\Delta$ we employ the sampling algorithm by \citet{legramanati_etal2022} to obtain a draw for the estimated group membership. For each node $j=1,\dots,M$ it performs the following steps:
\begin{enumerate}
    \item Remove $j$ from the network.
    \item Check whether group $z_j$ which previously included $j$ is now empty. If necessary, eliminate it and relabel all cluster indicators such that each cluster $h=1,\dots,H$ contains at least one node.
    \item Sample $z_j$ from the conditional posterior:
    \[
    p(z_j=h|\bm z_{-j}, \bm \Delta) \propto
    \begin{cases}
        \mathcal{W}_{M,H^-}(n_h^- - \sigma) \prod_{k=1}^{H^-} \frac{\mathcal{B}(a_\pi+m_{hk}^- + r_{jk}, b_\pi + \Bar{m}_{hk}^- + \Bar{r}_{jk})}{\mathcal{B}(a_\pi+m_{hk}^-,b_\pi+\Bar{m}_{hk}^-)} & \text{for } h\leq H^-, \\
        \mathcal{W}_{M,H^-+1} \prod_{k=1}^{H^-} \frac{\mathcal{B}(a_\pi+r_{jk}, b_\pi + \Bar{r}_{jk})}{\mathcal{B}(a_\pi,b_\pi)} & \text{for } h = H^-+1,
    \end{cases}
    \]
    where $\bm z_{-j}$ are the cluster membership indicators, $n_h$ is the size of cluster $h$, $m_{hk}^- (\Bar{m}_{hk}^-)$ are the number of links (non-existing links) between clusters $h$ and $k$ and $H^-$ is the number of non-empty clusters, all after removing $j$. We denote the number of links (non-existing links) between node $j$ and the $k$-th cluster as $r_{jk} (\Bar{r}_{jk})$. $\mathcal{W}_{M,H^-}$ and $\mathcal{W}_{M,H^-+1}$ are weights and $\sigma$ is a discounting factor that controls how quickly the sampler opens up a new cluster. All of the latter are determined by the specification of the Gibbs-type prior.
\end{enumerate}

\section{Data Appendix}\label{app:data}
An overview of the variables used in both models and their transformations can be found in table \ref{table:fred}.
\begin{landscape}
    \begin{table}[h!]
        \centering
        \begin{tabular}{c l c l c c}
            \toprule \toprule
            \textbf{FRED Code} & \textbf{Variable Name} & \textbf{tcode} & \textbf{Group} & \textbf{Medium} & \textbf{Large} \\ 
            \midrule
            GDPC1 & Real Gross Domestic Product & 5 & NIPA & x & x \\ 
            PCECC96 & Real Personal Consumption Expenditures & 5 & NIPA & x & x \\ 
            GPDIC1 & Real Private Domestic Investment & 5 & NIPA &  x & x \\ 
            PRFIx & Real Private Fixed Investment & 5 & NIPA & x & x \\ 
            INDPRO & Industrial Production Index & 5 & IP &  & x \\ 
            CUMFNS & Capacity Utilization: Manufacturing & 2 & IP &  & x \\ 
            UNRATE & Civilian Unemployment Rate & 2 & Labor Market & x & x \\ 
            SRVPRD & All Employees: Services & 5 & Labor Market & & x \\ 
            CE16OV & Civilian Employment Level & 5 & Labor Market & & x \\ 
            AWHMAN & Avg Weekly Hours of Production \& Nonsupervisory & 1 & Labor Market & & x\\ & Employees: Manufacturing & & & & \\
            CES3000000008x & Avg Hourly Earnings : Manufacturing & 5 & Earnings & x & x \\ 
            CPIAUCSL & Consumer Price Index for All Urban Consumers: & 6 & Prices & x & x\\ & All Items & & & & \\
            PCECTPI & Personal Consumption Expenditures: Chain-type & 6 & Prices  & x & x \\ & Price Index & & & & \\
            GDPCTPI & Gross Domestic Product: Chain-type Price Index & 6 & Prices & & x \\
            GPDICTPI & Gross Private Domestic Investment: Chain-type & 6 & Prices & & x \\ & Price Index & & & & \\
            FEDFUNDS & Effective Federal Funds Rate & 2 & Interest Rates & & x \\ 
            GS1 & 1-Year Treasury Rate & 2 & Interest rates & x & \\
            BAA10YM & Moody's Seasoned Baa Corporate Bond Yield & 2 & Interest Rates &  & x\\ & Rel. to 10-Year Treasury Constant Maturity & & & & \\ 
            M2REAL & Real M2 Money Stock & 5 & Money \& Credit & & x \\ 
            S\&P 500 & S\&P 500 Index & 5 & Stock Market & x & x \\
            \bottomrule \bottomrule
        \end{tabular}
        \caption{Macroeconomic variables alongside their assigned groups (as per \citet{mccracken_ng2016_jbes}) and their FRED identifier and tcode. tcode refers to the following transformations: 1) None 2) $\Delta x_t$ 5) $\Delta \text{log}(x_t)$ 6) $\Delta^2 \text{log}(x_t)$} 
        \label{table:fred}
    \end{table}
\end{landscape}

\end{document}